# FATE OF PRIORITY PHARMACEUTICALS AND THEIR MAIN METABOLITES AND TRANSFORMATION PRODUCTS IN MICROALGAE-BASED WASTEWATER TREATMENT SYSTEMS

María Jesús García-Galán[1], Larissa Arashiro[1], Lúcia H.M.L.M. Santos[2,3], Sara Insa[2,3], Sara Rodríguez-Mozaz[2,3], Damià Barceló[2,3,4], Ivet Ferrer[1], Marianna Garfi[1]

1. GEMMA, Group of Environmental Engineering and Microbiology, Department of Civil and Environmental Engineering, Universitat Politècnica de Catalunya-BarcelonaTech, c/ Jordi Girona 1-3, Building D1, E-08034 Barcelona, Spain
2. Catalan Institute for Water Research (ICRA), C. Emili Grahit 101, 17003 Girona, Spain
3. Universitat de Girona, Girona, Spain
4. Water and Soil Quality Research Group, Department of Environmental Chemistry, IDAEA-CSIC, c/ Jordi Girona 18-26, 08034 Barcelona, Spain

**GRAPHICAL ABSTRACT**

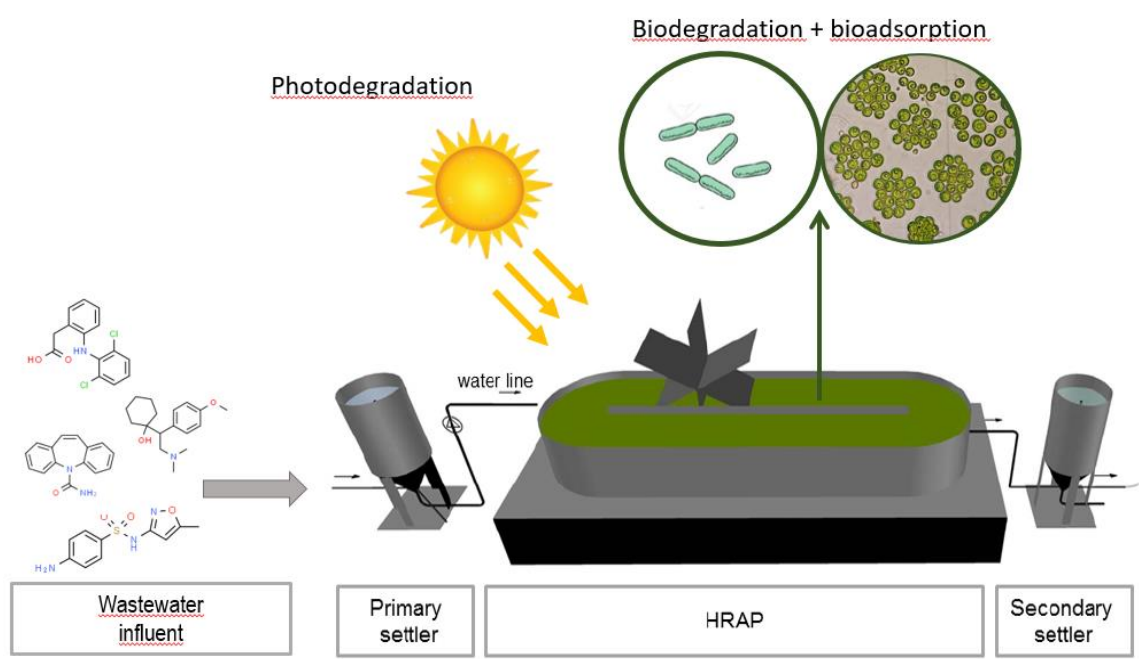

# ABSTRACT

The present study evaluates the removal capacity of two high rate algae ponds (HRAPs) to eliminate 12 pharmaceuticals (PhACs) and 26 of their corresponding main metabolites and transformation products. The efficiency of these ponds, operating with and without primary treatment, was compared in order to study their capacity under the best performance conditions (highest solar irradiance). Concentrations of all the target compounds were determined in both water and biomass samples. Removal rates ranged from moderate (40-60%) to high (>60%) for most of them, with the exception of the psychiatric drugs carbamazepine, the β-blocking agent metoprolol and its metabolite, metoprolol acid. O-desmethylvenlafaxine, despite its very low biodegradability in conventional wastewater treatment plants, was removed to certain extent (13-39%) Biomass concentrations suggested that bioadsorption/bioaccumulation to microalgae biomass was decisive regarding the elimination of some non-biodegradable compounds such as venlafaxine and its main metabolites. HRAP treatment with and without primary treatment did not yield significant differences in terms of PhACs removal efficiency. The implementation of HRAPs as secondary treatment is a viable alternative to CAS in terms of overall wastewater treatment (including organic micropollutants), with generally higher removal performances and implying a green, low-cost and more sustainable technology.

**KEYWORDS:** high rate algal pond, green treatment, low-cost treatment, bioaccumulation factor, photodegradation, byproducts, biomass

**INTRODUCTION**

Our increasing dependence on pharmaceutical compounds (PhACs) has led to their widespread presence in all kind of environmental compartments, and at global level. PhACs are bioactive molecules that can remain as such even after being metabolized in the organism. They are generally highly resilient to biodegradation and so their removal during conventional wastewater treatment (based on activated sludge, (CAS)), is usually incomplete. Indeed, during the last decade, several studies have demonstrated that CAS treatment does not completely eliminate the vast majority of organic micropollutants, including PhACs (Archana et al., 2017; Bradley et al., 2017; García-Galán et al., 2011; Gros et al., 2007; Kostich et al., 2014; Li et al., 2019; Madikizela et al., 2017). However, CAS treatment does comply with the Council Directive 91/271/EEC requirements (European Union, 1991) regarding quality of wastewater effluents prior to their discharge in the environment (chemical oxygen demand (COD), biological oxygen demand (BOD), nitrogen, phosphorus and total suspended solids (TSS)). Highly prescribed/used compounds such as the anti-inflammatory diclofenac (DCF), the antiepileptic carbamazepine (CBZ) or the antibiotic sulfamethoxazole (SMX) are not completely removed by CAS WWTPs and they are regularly detected in environmental waters (García-Galán et al., 2011; Gros et al., 2010). Tertiary treatments such as advanced oxidation processes (AOPs) (i.e. UV, photo-fenton, ozonation), ultrafiltration or nanofiltration are then required to eliminate these pollutants more efficiently and obtain an effluent of much higher quality (Mamo et al., 2018; Taheran et al., 2016). These processes, however, are usually economically unfeasible and, in the case of AOPs, they put forward a new problem regarding the potential ecotoxicity of the different transformation products (TPs) generated during the process (Fatta-Kassinos et al., 2011; García-Galán et al., 2016; Hübner et al., 2015). Nevertheless, the aforementioned low removal efficiency results in the regular entrance of these compounds onto the aquatic environment via discharge of wastewater effluents, constituting an ecotoxicological threat to non-target aquatic organisms directly or indirectly exposed to these substances (bioaccumulation and/or biomagnification) (Garcia-Galan et al., 2017; Langdon et al., 2010; Ruhí et al., 2016). Furthermore, and given the magnitude of the problem, scientific concern has also moved from the conventional monitoring studies on the presence of these PhACs and other pollutants in WWTP effluents or surface waters, to consider the presence of not only the parent drugs but also their metabolites and TPs and the ecotoxicological effects of the mixtures (Aristi et al., 2015; García-Galán et al., 2016).

Taking all this into consideration, the search of alternative and more efficient treatments has become imperative. In particular, green, low-cost systems such as those based on microalgae have been investigated with great interest due to their high capability to remove nutrients, heavy metals and bacteria (García et al., 2006; Muñoz and Guieysse, 2006). Open systems (usually high-rate algal ponds (HRAPs)) are the most commonly used, due to its versatility, easy operation and low maintenance costs. HRAPs are shallow raceway reactors in which microalgae and bacteria grow in symbiosis. Microalgae can grow in low quality water such as wastewater effluents, using up the nutrients they still contained, and releasing oxygen during photosynthesis. Concomitantly, the oxygen produced supports the respiration of heterotrophic bacteria, which are able to degrade organic matter. This symbiosis leads to the production of clean water as a by-product, and at the same time, to the growth and production of algal biomass that can be further stabilized, processed and converted to bioenergy (biodiesel and/or biogas) or biofertilizer (Benemann et al., 2013; Chisti, 2013; Thomas et al., 2016). Further advantages have been recently presented, such as the production and intracellular accumulation, under specific growing conditions, of different added-value products such as glycogen, bioplastics or terpenoids, that can also be recovered from microalgae (Arias et al., 2018; Khetkorn et al., 2017, 2016; Uggetti et al., 2018). Taking all this into consideration, HRAP treatment could be considered as one of the most environmentally and economical favorable due to the reduced energy requirements and low operation and maintenance costs. Solar irradiation is therefore the main limiting factor regarding optimal growth rates and the biodegradative activity of microalgae (Nurdogan and Oswald, 1995).

Despite the high efficiency of microalgae in removing nutrients and organic matter from wastewater, studies on their potential to eliminate organic micropollutants such as PhACs are still very scarce. Most importantly, metabolites or TPs of the original pharmaceuticals are very seldom addressed (García-Galán et al., 2018; Jaén-Gil et al., 2018; Matamoros et al., 2015). In addition, data on biodegradation of PhACs by microalgae is usually based on lab-scale experiments carried out under sterile and controlled conditions, which are environmentally unrealistic (de Wilt et al., 2016; Ding et al., 2017). As happens with CAS treatments, HRAPs are not designed specifically to remove PhACs or other organic micropollutants, but their longer hydraulic residence times (HRTs) (usually several days vs hours of CAS), could favour the biodegradation of those contaminants with higher half-lives (and slow kinetics biodegradation). Other advantages to be highlighted are their high surface-area-to-volume ratio and subsequent higher sunlight exposures (enhancing the photodegradation pathways), the coexistence of autotrophic and heterotrophic microorganisms that improves the biomass productivity (and removal by adsorption to biomass), as well as biodegradation by different phototrophic, chemoorganotrophic and chemolithotrophic metabolic pathways (García et al., 2006; Nurdogan and Oswald, 1995).

Daily changes in pH, dissolved oxygen (DO) and redox conditions could also improve the removal mechanisms of PhACs in these systems (Norvill et al., 2017).

The present study aims to evaluate the capacity of HRAPs to attenuate the concentration of 12 PhACs and 26 of their main metabolites and/or TPs, and to elucidate the removal mechanisms involved. Two HRAPs systems with different configurations (with and without primary treatment) were set up and compared during the beginning of summer, coinciding with the highest solar irradiation rates.

## 2. MATERIALS AND METHODS

### 2.1. HRAPs pilot plant description

Samples were taken from a microalgae-based pilot plant installed outdoors at the laboratory of the GEMMA research group (Universitat Politècnica de Catalunya–BarcelonaTech, Spain) and previously defined by Matamoros et al. (2015). Briefly, it consisted of two parallel treatment lines formed each by a primary settler, a HRAP and a settler or clarifier to separate the biomass produced from the treated effluent (see Figure 1). Each HRAP had a volume of 470 L (surface area of 1.54 $m^2$, 0.3 m depth) and was equipped with a paddle wheel (working at 5 rpm approximately) in the middle of the pond to ensure the correct mixing of the liquor. Both systems were fed with domestic wastewater from the surrounding neighborhoods to the campus (residential areas), which was directly pumped from a municipal sewer to a homogenization tank (1.2 $m^3$), which was continuously stirred to avoid solids sedimentation.

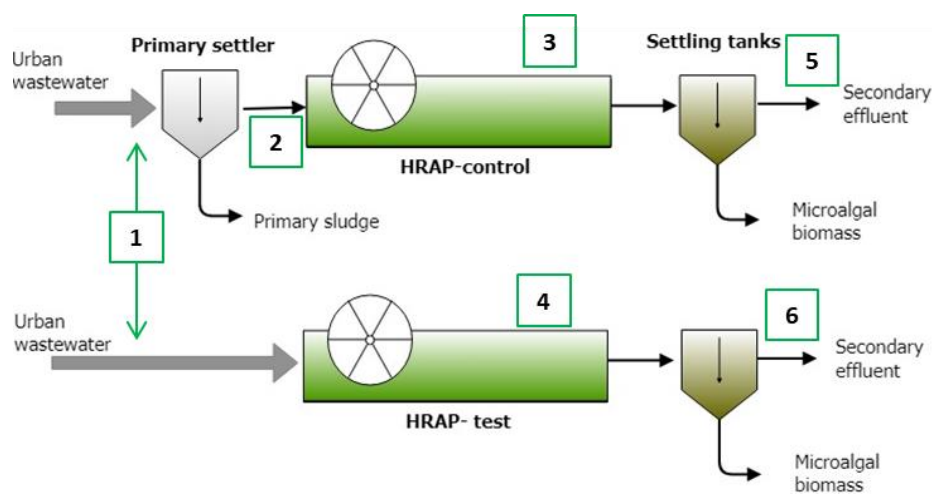

**Figure 1**. Diagram of the two HRAPs working in parallel and sampling points: 1. Raw wastewater/ HRAP-test influent; 2: HRAP-control influent (after primary settler); 3: HRAP-control mixed liquor; 4: HRAP-test mixed liquor; 5: HRAP-control effluent; 6: HRAP-test effluent.

In order to evaluate the influence of primary treatment in the overall removal of the target PhACs, the primary settler of one of the lines was inoperative, so that the efficiency of both configurations could be compared. The standard configuration of the HRAP (including the primary settler) was considered as the reference/control HRAP (HRAP-control), whereas the second HRAP without that primary treatment was named HRAP-test (Figure 1). The systems were operating in continuous since November 2016 at a HRT of 4.5 days. Further details on the operation of these systems are described elsewhere by Arashiro et al., (2019).

## 2.2. Sampling strategy

Sampling was carried out during two consecutive weeks in June 2017. Grab samples for the influent and effluent water of both ponds were taken daily from Monday to Friday from the 19$_{th}$ of June till the 30$_{th}$ of June 2017 (n=40 samples). For chemical characterization of the water, samples were taken in PVC bottles and directly analyzed in the laboratory. For PhACs analysis (environmental/real levels), samples were collected and immediately filtered through 0.45 μm PVDF membrane filters (Millipore, USA) and frozen until analysis (amber glass bottles).

Biomass samples were taken twice a week from each HRAP configuration (n=8 samples) also in PVC samplers. They were immediately centrifuged at 4200 rpm for 10 min (Unicen 21, Ortoalresa, Spain), frozen at −80°C overnight in an ultra-freezer (Arctiko, Denmark) and finally freeze-dried for 24 h (−110 °C, 0.049 hPa) (Scanvac, Denmark). The biomass samples were then stored at -21°C until analysis.

### 2.3. Chemicals and reagents

High purity standards (>99%) of the pharmaceuticals acetaminophen (ACM), sulfadiazine (SDZ), sulfamethazine (SMZ), sulfamethoxazole (SMX), venlafaxine (VFX), diazepam (DZP), carbamazepine (CBZ), diclofenac (sodium salt) (DCF), fluoxetine (FXT), ibuprofen (IBF), metoprolol (MTP), metronidazole (MTZ), acetylsalicylic acid (ASA) and the metabolites norfluoxetine (desFXT) and 2-OH-carbamazepine (CBZ-OH) were purchased from Sigma-Aldrich (St. Louis. MO. USA). High purity standards for the metabolites ibuprofen carboxilyc acid (cbz-IBF), 1-OH- ibuprofen (1-IBF-OH), rac-2-OH-ibuprofen (2-IBF-OH), 4'-OH-diclofenac (DCF-OH), diclofenac amide (adDCF), diclofenac acyl-B-D-glucuronide (gluDCF), acridone (ACRO), D,L-N-desmethyl venlafaxine (N-desVFX), D,L-O-desmethyl venlafaxine (O-desVFX), D,L-N,N-Didesmethyl venlafaxine, D,L-N,N-Didesmethyl-O-desmethyl venlafaxine, N,O-Didesmethyl venlafaxine, $N_4$-acetylsulfadiazine (acSDZ), $N_4$-acetylsulfamethazine (acSMZ), $N_4$-acetylsulfamethoxazol (acSMX), 4-nitro-sulfamethoxazole (4-nitro-SMX), desmethyldiazepam (desDZP), 3-OH-acetaminophen (ACM-OH), metoprolol acid (MTPA), α-OH-metoprolol (α-OH-HMTP), O-desmethylmetoprolol (O-DMTP), metronidazole-OH (MTZ-OH) and 10,11-epoxy carbamazepine (epoCBZ) were purchased from TRC (Toronto Research Chemicals Inc.. Ontario. Canada). Isotopically labelled compounds, used as internal standards were purchased from Sigma-Aldrich (fluoxetine-$d_5$, ibuprofen-$d_3$), TRC *(*diclofenac-$d_4$, 4'-OH-diclofenac-$d_4$, sulfamethoxazole-$d_4$, N,L-O-desmethylvenlafaxine-$d_4$ and acetaminophen-$d_4$), Cerilliant (Texas, U.S.A.) (diazepam-$d_5$) and from CDN isotopes (Quebec, Canada) (carbamazepine-$d_{10}$ and venlafaxine-$d_6$). Stock standard solutions for each of the compounds were prepared in MeOH at 1mg mL$^{-1}$ and stored in the dark at −18 °C. Standard solutions of the mixtures of all compounds were made at appropriate concentrations and used to prepare the aqueous calibration curve and also to perform the recovery studies. Similarly, stock standard solutions for the internal standards were prepared. Aqueous standard solutions always contained <0.1% of MeOH.

## 2.4. Analytical methodologies

### 2.4.1. Samples characterization

Both influent and effluent water samples, together with the mixed liquor within the ponds (before the secondary settler) were analyzed on the following parameters: dissolved oxygen (DO) and temperature (EcoScan DO 6, ThermoFisher Scientific, USA) (daily); pH (Crison 506, Spain) and turbidity (Hanna HI 93703, USA), (three times per week); total suspended solids (TSS), volatile suspended solids (VSS), chlorophyll-*a*, according to Standard Methods (APHA-AWWA-WEF, 2012), $NH_4^+$-N according to Solórzano method (Solórzano, 1969) and $NO_2^-$-N, $NO_3^-$-N and $PO_4^{3-}$-P by ion chromatography (ICS-1000, Dionex Corporation, USA), twice a week). Ion chromatography was performed in isocratic mode with carbonate-based eluents at a temperature of 30°C and a flow of 1 mL min$^{-1}$. Limits of detection (LOD) obtained were 0.9 mg/L of $NO_2^-$-N, 1.12 mg/L of $NO_3^-$-N, and 0.8 mg/L of $PO_4^{3-}$-P; alkalinity, total and soluble chemical oxygen demand (COD and sCOD) according to Standard Methods (APHA-AWWA-WEF, 2012), total carbon (TC) and total nitrogen (TN) (multi N/C 2100S, Analytik Jena, Germany) (once a week). All the analyses were done in triplicate and results were given as average values. Mixed liquor samples were regularly examined under an optic microscope (Motic, China) for qualitative evaluation of microalgae populations, employing taxonomic books and databases for their identification (Bourrelly, 1990; Palmer, 1962).

*Biomass productivity.*

Average biomass productivity (gVSS $m^{-2} \cdot d^{-1}$) was calculated based on the VSS concentration in the HRAPs mixed liquor samples and following (Arashiro et al., 2019), using [1]:

$$Biomass\ productivity = \frac{VSS\ (Q - Q_E + Q_P)}{A} \quad [1]$$

where $VSS$ is the volatile suspended solids concentration of the HRAP mixed liquor (g VSS $L^{-1}$); $Q$ is the wastewater flow rate (L $d^{-1}$); $Q_E$ is the evaporation rate (L $d^{-1}$); $Q_P$ is the precipitation rate (L $d^{-1}$); and $A$ is the surface area of the HRAP (1.54 $m^2$). The evaporation rate was calculated using [2]:

$$Q_E = E_p\ A \quad [2]$$

where $E_p$ is the potential evaporation (mm $d^{-1}$), calculated using [3] (Fisher and Pringle III, 2013).

$$E_p = a\ \frac{T_a}{(T_a + 15)} (R + 50) \quad [3]$$

where $a$ is a dimensionless coefficient which varies depending on the sampling frequency (0.0133 for daily samples); $R$ is the average solar radiation in a day (MJ $m^{-2}$), and $T_a$ is the average air temperature (°C). Meteorological data were obtained from the net of local weather stations in Barcelona and metropolitan area (*www.meteo.cat*).

### 2.4.2. UHPLC-MS/MS analysis of the target compounds

*Biomass samples*

Microalgae biomass samples were pre-treated and readied for UHPLC-MS/MS analysis following the method by Santos et al. (2019), which is briefly detailed in SI.

*Water samples*

The different PhACs, metabolites and TPs were analyzed using a methodology adapted from García-Galán et al., (2016), and detailed in SI. Detailed data on method validation for water and biomass samples are given in SI. For both types of analysis, water and solid matrices, the performance of the methodology in terms on limits of detection and quantification of the method, recoveries of the biomass samples and matrix effects are given in SI.

## 3. RESULTS AND DISCUSSION

### 3.1. Overall performance of the HRAPs

Table 1 summarizes the operation conditions of both HRAPs, with no precipitation events and regular temperature of 20.7 ºC.

**Table 1.** On-site parameters (average values) measured for both HRAPs during the sampling campaign

|  | Flow rate (L/d) | HRT (d) | DO (mg/L) | Temperature (ºC) | Turbidity (NTU) | pH |
|---|---|---|---|---|---|---|
| **HRAP-C** | 111.7 ± 6 | 4,5 | 8.2 ± 1 | 20.8 ± 2.4 | 75.9 ± 63 | 8.3 ± 0.6 |
| **HRAP-T** | 104.2 ± 6.2 | 4,2 | 7.1 ± 0.9 | 20.7 ± 2.4 | 110.8 ± 73.2 | 8.2 ± 0.1 |

As expected, the absence of primary treatment led to higher TSS concentrations in HRAP-test (29% higher in average) due to the greater inorganic solids concentrations entering this pond directly from the sewage system (Figure 2). VSS was also a 31% higher in HRAP-test. Regarding the physical-chemical quality parameters, both HRAP configurations presented similar nutrients and organic matter removal efficiencies (Table 2). The removal of $NH_4^+$-N was > 95% in both cases. Generally, $NH_4^+$-N is the preferential form of nitrogen uptake for most microalgae species, followed by $NO_3^-$-N (Maestrini, 1982; Oliver and Ganf, 2002;

Ruiz-Marin et al., 2010). Furthermore, the high photosynthetic activity during summer increases the pH and favours $NH_4^+$-N volatilization, and also accelerates its partial nitrification to $NO_3^-$; indeed, despite the concentrations of $NO_3^-$ and $NO_2^-$ in the influent were very low, these were barely removed or not eliminated. Nitrification of the influent N content (mostly $NH_4^+$) into $NO_3^-$-N and $NO_2^-$-N could explain those negative removals (de Godos et al., 2016; Van Den Hende et al., 2016). All in all, the TN removal efficiencies were positive but lower than those of $NH_4^+$-N (between 72% and 61% in HRAP-control and HRAP-test, respectively). Considering the maximum discharge values for TN registered in the Council Directive 91/271/EEC, we observe that the boundary limit of 15 mg $L^{-1}$ is surpassed in both HRAPs. However, it should be noted that this value is given for systems treating wastewater for PE of 10 000-100 000, which is far bigger than the treatment capacity of the pilot HRAPs studied in this work.

Average $COD_t$ and $COD_s$ removal efficiencies were very similar for both ponds, and are in accordance with previous studies under similar operational conditions (Young et al., 2017; Sutherland et al., 2014). Effluent values comply with the Council Directive aforementioned.

**Table 2.** Removal rates obtained for the different parameters evaluated in the wastewater samples during the sampling campaign

|  | Units | HRAP-CONTROL | | | HRAP-TEST | | |
| --- | --- | --- | --- | --- | --- | --- | --- |
|  |  | Influent | Effluent | R% | Influent | Effluent | R% |
| **$COD_t$** | mg $L^{-1}$ | 199 | 78 | **60.8** | 243 | 93 | **61.7** |
| **$COD_s$** | mg $L^{-1}$ | 63 | 44 | **30.2** | 57 | 40 | **29.8** |
| **$NH_4^+$-N** | mg $L^{-1}$ | 24.9 | 1.2 | **95.2** | 24.1 | 0.6 | **97.5** |
| **$NO_3^-$-N** | mg $L^{-1}$ | 1.1 | 11.4 | - | 1.0 | 8.5 | - |
| **$NO_2^-$-N** | mg $L^{-1}$ | 4.2 | 3.4 | **19.0** | 2.4 | 3.0 | - |
| **TN** | mg $L^{-1}$ | 79.1 | 22.4 | **71.7** | 86.1 | 33.8 | **60.7** |
| $PO_4^{3-}$-P | mg L$^{-1}$ | 3.4 | 2.2 | **35.3** | 4.1 | 7.5 | - |
| $SO_4^{2-}$-S | mg L$^{-1}$ | 44.4 | 53 | - | 39.9 | 51.2 | - |

Regarding the biomass productivity, focusing in the two weeks of sampling (19th-30th June), a higher productivity in HRAP-control than in HRAP-test was observed, especially during the first week (14.6 gVSS m$^{-2}$·d$^{-1}$ vs 5 gVSS m$^{-2}$·d$^{-1}$) (Figure 3). Solar irradiance was very stable and similar during those days (see Figure S1 of SI). As mentioned, the higher TSS observed in HRAP-test blocked the light penetration in the pond, reducing the photosynthetic activity and the biomass growth. The highest productivity was achieved by the end of the sampling campaing, with values of 21.7 gVSS m$^{-2}$·d$^{-1}$ in HRAP-control and 15.7 gVSS m$^{-2}$·d$^{-1}$ in HRAP-test. These results are similar to those obtained in previous

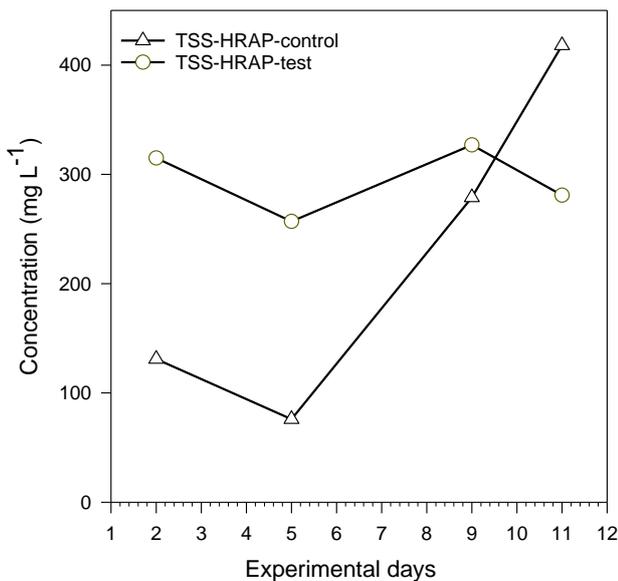
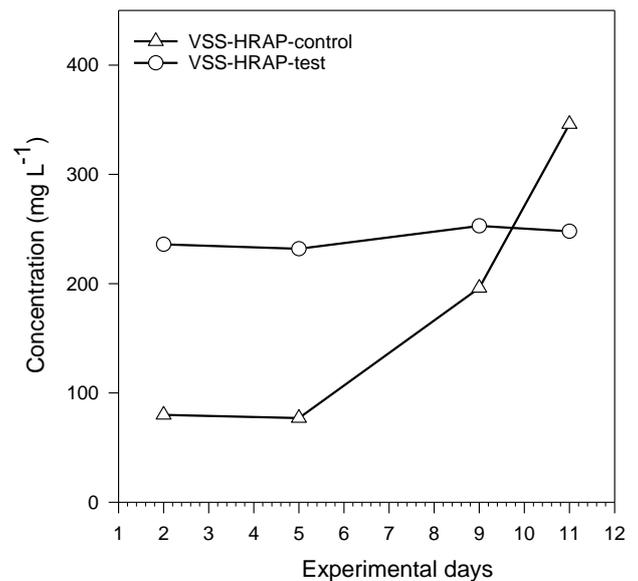

works in HRAPs (Norvill et al., 2017; Sutherland et al., 2014).

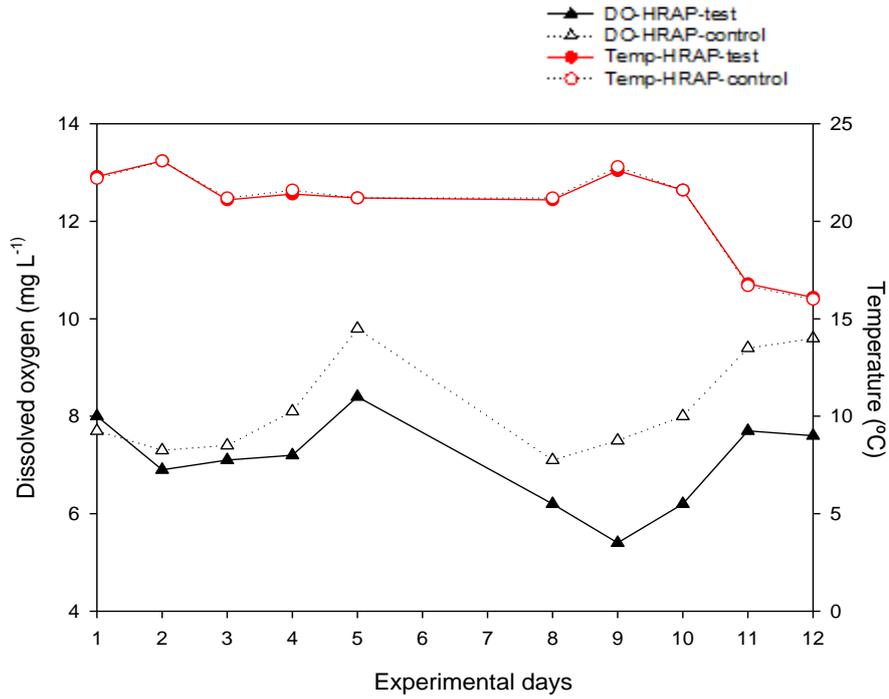

**Figure 2.** Total suspended solids (TSS) and volatile suspended solids (VSS), dissolved oxygen and temperature measured in both HRAPs during the month of June.

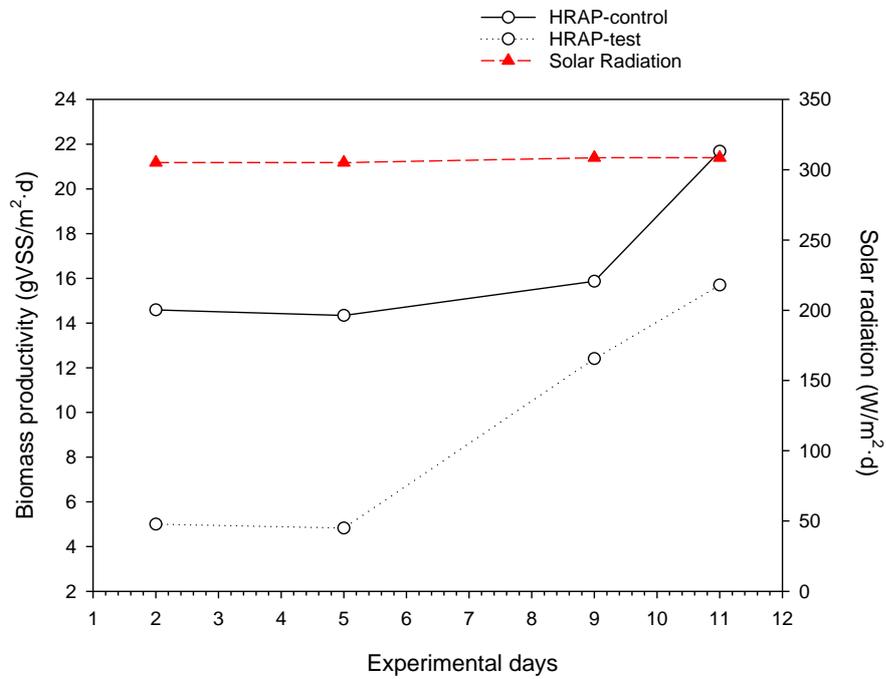

**Figure 3**. Biomass productivity obtained in both HRAPs during the sampling campaign

**3.2. Occurrence of PhACs in raw urban wastewater and primary treatment**

Nine compounds out of the 12 parent compounds and 15 out of the 26 metabolites and/or TPs were detected in influent wastewaters, with no significant differences between the concentrations observed before and after primary treatment (Table 3). The highest concentrations corresponded to IBF, a non-steroidal anti-inflammatory drug (NSAID), and one of its main human metabolites, 2-hydroxy-ibuprofen (2-OH-IBF), with levels in the range of 330-23811 ng $L^{-1}$ for IBF and 15324-147755 ng $L^{-1}$ for the metabolite. IBF is one of the most consumed anti-inflammatories in Europe, with estimated yearly consumptions in Spain of 250 tons (Ortiz de García et al., 2013). Another main metabolite, carboxy-ibuprofen (cbx-IBF) was found in only a few samples, at a concentration ranging between 16697 ng $L^{-1}$ – 44211 ng $L^{-1}$, one order of magnitude lower than levels detected in previous studies (Ferrando-Climent et al., 2012). This may be indicative of biodegradation events before reaching the HRAPs, as demonstrated recently by Jelic et al. for different PhACs in sewer pipes before reaching WWTPs (Jelic et al., 2015). Another NSAID, DCF, was detected in all the influent wastewater samples at concentrations ranging from 271 to 2126 ng $L^{-1}$. Its hydroxilated moieties, 4-hydroxy-diclofenac and 5-hydroxydiclofenac (4-OH-DCF y 5-OH-DCF), were also frequently detected at significant levels (see Table 3). DCF is also one of the most widely prescribed anti-inflammatory and its incomplete removal during CAS treatment has led to its widespread presence in the environment [1] (Huerta et al., 2015; Oaks et al., 2004) . Acetylsalicylic acid (AcSAc), also a NSAID, was detected all days of sampling in the influent wastewater, but at very variable concentrations that ranged between 166 ng $L^{-1}$ and 5428 ng $L^{-1}$. This variability was also observed in previous works (Gracia-Lor et al., 2012; Gros et al., 2012). Acetaminophen (ACM) is also an analgesic frequently detected in WWTPs influents at levels similar to IBF (Gros et al., 2012; Mamo et al., 2018); in the present study, however, it was detected at a maximum concentration of 7800 ng $L^{-1}$.

Regarding antibiotics, SMX and its acetylated metabolite, acSMX, were detected in all the influent samples (control and test), at concentrations ranging between 70 and 900 ng L$^{-1}$ for SMX and much higher, 654 ng L$^{-1}$ to 10188 ng L$^{-1}$ ,for the metabolite, agreeing with the results of previous studies in which both compounds were detected in domestic wastewaters (García-Galán et al., 2016; Mamo et al., 2018; Rodriguez-Mozaz et al., 2015). This antibiotic is considered a ubiquitous pollutant in natural systems. Furthermore, among the family of sulfonamide antibiotics, SMX has been identified as the main responsible for the increase of sulfonamide resistance genes in WWTPs (Guo et al., 2017).

The anti-depressant venlafaxine (VFX) and its two main metabolites, O-desmethylvenlafaxine (O-desVFX) and N-desmethylvenlafaxine (N-desVFX), were present at average concentrations ranging between 664-702 ng L$^{-1}$ for VFX, 738-817 ng L$^{-1}$ for O-desVFX and 108-115 ng L$^{-1}$ for N-desVFX. The occurrence of three other minor VFX metabolites was also evaluated: N,N-didesmethylvenlafaxine (N,N-ddVFX) and N,O-didesmethylvenlafaxine (N,O-ddVFX), which were detected at concentrations < 200 ng L$^{-1}$; and N,N-Didesmethyl-O-desmethyl venlafaxine (N,N-dd-O-dVFX), which remained below LOD. VFX is widely used in the US and Europe, and its presence in wastewater has been widely demonstrateds ( García-Galán et al., 2016; Mamo et al., 2018). VFX undergoes metabolism in the liver, converting to its major metabolite, O-desVFX, an active compound which is itself also commercialized in Spain since 2012 (AEMPS, 2012). This metabolite is ubiquitous in wastewaters and surface waters and usually at concentrations two and three folds higher than those of VFX (García-Galán et al., 2016; Mamo et al., 2018). The antiepileptic carbamazepine (CBZ) and one of its main metabolites, 2-hydroxy-carbamazepine (2-OH-CBZ), were present in all the influent samples but at concentrations < 33 ng L$^{-1}$, values slightly lower than those found in previous studies (Dolar et al., 2012; García-Galán et al., 2016).

Metoprolol (MTP) was detected only in the influent samples of the HRAP-control, at concentrations ranging between 40 and 210 ng L$^{-1}$, whereas one of its main metabolites, metoprolol acid (MTPA), was detected in all of them, at similar concentrations. These values are lower than those previously reported (Mamo et al., 2018; Rubirola et al., 2014)

Table 3. Concentrations of the different target compounds studied in the influent wastewater (n=10) of the ponds (ng L$^{-1}$) and in the biomass (ng g$^{-1}$ dw). FD: frequency of detection; (-): not detected; <LOQ: below the method limit of quantification

| | | HRAP-control | | | | | | | |
|---|---|---|---|---|---|---|---|---|---|
| | | WATER | | | | BIOMASS | | | |
| | COMPOUND | AVERAGE (ng L$^{-1}$) | Std | Concentration range (ng L$^{-1}$) | FD% | AVERAGE (ng g$^{-1}$) | Std | Concentration range (ng L$^{-1}$) | FD% |
| Antibiotics | Sulfamethoxazole | 383.36 | 295.0 | 69.4-902.5 | 60 | 38.16 | 15.3 | 27.4-55.7 | 100 |
| | N$^4$–acetylsulfamethoxazole | 2367.64 | 1827.4 | 653.8-10187.7 | 60 | - | - | - | - |
| | Nitrosulfamethoxazole | - | - | - | - | 48.12 | 7.0 | 39.8-56.7 | 100 |
| | Sulfamethazine | 95.24 | - | - | 10 | - | - | - | 100 |
| | N$^4$-Acetylsulfamethazine | - | - | - | - | - | - | - | 100 |
| | Sulfadiazine | - | - | - | - | - | - | - | 100 |
| | N$^4$-Acetylsulfadiazine | 1704.53 | - | - | 10 | - | - | - | 100 |
| | Metronidazole | 369.83 | 676.9 | 23.5-1742.4 | 60 | - | - | - | 100 |
| | OH-metronidazole | - | - | - | - | 548.44 | 140.2 | 423.7-739.7 | 100 |
| Psychiatric drugs | Venlafaxine | 625.84 | 230.8 | 382.4-1005.9 | 100 | 524.34 | 157.2 | 355.6-766 | 100 |
| | N–Desmethylvenlafaxine | 108.33 | 88.4 | 38.8-292.3 | 100 | 379.80 | 79.2 | 297.3-461.7 | 100 |
| | O–Desmethylvenlafaxine | 654.42 | 342.2 | 374.2-840.8 | 100 | 125.57 | 34.8 | 91.4-176.4 | 100 |
| | N,N-Didesmethylvenlafaxine | 67.51 | 33.3 | 44-91.1 | 20 | 112.78 | 18.3 | 93.5-130 | 75 |
| | N,N-Didesmethy-O-desmethyllvenlafaxine | - | - | - | - | - | - | - | - |
| | NO-desmethylvenlafaxine | 95.96 | 44.4 | 49.3-113.1 | 60 | 100.68 | 22.0 | 76.6-127 | 100 |
| | Fluoxetine | - | - | - | - | 113.73 | 25.5 | 86.2-140.2 | 100 |
| | Norfluoxetine | 15.16 | 2.1 | 13,7-16,6 | 20 | 69.84 | 16.1 | 58.4-81.3 | 75 |
| | Diazepam | - | - | - | - | 6.13 | 0.4 | 5.8-6.4 | 50 |
| | Desmethyldiazepam | 2.85 | 2.5 | - | 30 | - | - | - | - |
| | Carbamazepine | 20.95 | 10.8 | 9.6-42.4 | 100 | 24.94 | 8.1 | 18.7-36.5 | 100 |

| | Compound | AVERAGE (ng L⁻¹) | Std | Concentration range (ng L⁻¹) | FD% | AVERAGE (ng g⁻¹) | Std | Concentration range (ng L⁻¹) | FD% |
|---|---|---|---|---|---|---|---|---|---|
| | Epoxy– Carbamazepine | - | - | - | - | 0.98 | - | 0.98 | 50 |
| | 2–OH-carbamazepine | 32.09 | 20.9 | 11.1-75.7 | 100 | 2.86 | - | 2.86 | 25 |
| | Acridone | - | - | - | - | 52.03 | 19.4 | 29.1-76.1 | 100 |
| Analgesics/ | Acetylsalicylic acid | 845.44 | 462.7 | 272.6-4688.3 | 100 | - | - | - | <LOD |
| | Diclofenac | 793.09 | 638.4 | 270.8-2117.8 | 100 | 88.58 | 99.5 | 6.9-216.1 | 75 |
| | 4-OH-Diclofenac | 241.22 | 79.3 | 125.9-403.1 | 90 | 8.07 | - | 8.07 | 25 |
| | 5-OH-Diclofenac | 856.52 | 456.4 | 380.75-1464.8 | 40 | - | - | - | - |
| | Diclofenac–glucuronide | - | - | - | - | - | - | - | - |
| | Diclofenac amide | - | - | - | - | - | - | - | - |
| | Acetaminophen | 1579.38 | 3045.5 | 156.4-7781.1 | 60 | 93.98 | - | 93.98 | 25 |
| | Acetaminophen–OH | - | - | - | - | - | - | - | - |
| | Ibuprofen | 4731.18 | 7029.3 | 713.3-23811-4 | 100 | 467.69 | 176.9 | 283.4-671.7 | 75 |
| | 1-OH-ibuprofen | 4819.75 | 3157.2 | 1813.1-9007.2 | 40 | 79.60 | 59.6 | 29.4-145.4 | 75 |
| | 2-OH-ibuprofen | 49102.13 | 32790.1 | 15324.5-112839.7 | 100 | 152.21 | 85.4 | 78.9-242.2 | 75 |
| | Carboxy-ibuprofen | 22670.13 | 12069.5 | 16697-44211 | 50 | 38.45 | 31.4 | 10.6-74.6 | 100 |
| β-blocking agents | Metoprolol | 92.98 | 67.2 | 39.3-209.6 | 50 | 92.87 | 25.0 | 48.9-108.8 | 100 |
| | Metoprolol–acid | 102.96 | 59.7 | 33.1-207.1 | 100 | - | - | - | - |
| | Alfa–Metoprolol–OH | - | - | - | - | 35.44 | 7.2 | 28.4-44.9 | 100 |
| | O–Desmethyl–Metoprolol | - | - | - | - | - | - | - | - |

| | | HRAP-test | | | | | | | |
|---|---|---|---|---|---|---|---|---|---|
| | | WATER | | | | BIOMASS | | | |
| | COMPOUND | AVERAGE (ng L⁻¹) | Std | Concentration range (ng L⁻¹) | FD% | AVERAGE (ng g⁻¹) | Std | Concentration range (ng L⁻¹) | FD% |
| Antibiotics | Sulfamethoxazole | 345.03 | 263.5 | 14.2-744.9 | 50 | 49.55 | 25.5 | 31.1-78.7 | 75 |

| | | | | | | | | |
|---|---|---|---|---|---|---|---|---|
| | N⁴–acetylsulfamethoxazole | 2520.15 | 1565.9 | 901.2-4971.2 | 60 | - | - | - | - |
| | Nitrosulfamethoxazole | - | - | - | - | - | - | - | - |
| | Sulfamethazine | 469.20 | 498.6 | 116.3-821.8 | 20 | 51.19 | 8.8 | 42.3-65.3 | 100 |
| | N⁴-Acetylsulfamethazine | - | - | - | - | - | - | - | - |
| | Sulfadiazine | - | - | - | - | - | - | - | - |
| | N⁴-Acetylsulfadiazine | 2733.00 | - | - | 10 | - | - | - | - |
| | Metronidazole | 429.31 | 677.1 | 39.5-1626.9 | 50 | - | - | - | - |
| | OH-metronidazole | - | - | - | - | - | - | - | - |
| Psychiatric drugs | Venlafaxine | 633.74 | 256.9 | 348.8-1038.9 | 100 | 329.03 | 75.5 | 232.3-439.3 | 100 |
| | N–Desmethylvenlafaxine | 115.09 | 99.0 | 38.4-329.2 | 100 | 252.04 | 26.9 | 216.8-287.4 | 100 |
| | O–Desmethylvenlafaxine | 577.37 | 308.4 | 134.7-1042.5 | 100 | 312.00 | 45.7 | 267.8-384.5 | 100 |
| | N,N-Didesmethylvenlafaxine | 60.13 | 37.5 | 37.9-103.5 | 30 | 124.38 | 17.0 | 100.2-146 | 100 |
| | N,N-Didesmethy-O-desmethyllvenlafaxine | - | - | - | - | <LOQ | - | - | <LOQ |
| | NO-desmethylvenlafaxine | 87.81 | 67.0 | 30.6-195.1 | 60 | 132.94 | 14.0 | 120.8-149.2 | 100 |
| | Fluoxetine | - | - | - | | 187.26 | 53.1 | 119.1-240.8 | 75 |
| | Norfluoxetine | 14.41 | 4.1 | 11.5-17.3 | 20 | 102.13 | 51.8 | 44-169.4 | 75 |
| | Diazepam | - | - | - | - | 9.03 | 1.0 | 7.9-10.4 | 100 |
| | Desmethyldiazepam | 2.63 | 0.9 | 2-3.3 | 20 | - | - | - | - |
| | Carbamazepine | 21.58 | 11.9 | 9.7-46.2 | 100 | 14.96 | 2.8 | 10.9-18.1 | 100 |
| | Epoxy– Carbamazepine | 5.34 | 0.6 | 4.9-5.7 | 20 | 7.69 | 1.7 | 6.5-8.9 | 50 |
| | 2–Hydroxycarbamazepine | 29.24 | 17.2 | 10.9-68.4 | 100 | 6.60 | 2.2 | 3.6-8.8 | 100 |
| | Acridone | - | - | - | | 77.76 | 15.6 | 61.1-97.5 | 100 |
| Analgesics/ antinflammatories | Acetylsalicylic acid | 1307.36 | 1580.3 | 165.8-5428.1 | 100 | - | - | <LOD | - |
| | Diclofenac | 784.87 | 560.7 | 288.1-2126.5 | 100 | 267.98 | 159.6 | 283.1-424.4 | 75 |
| | 4-OH-Diclofenac | 217.26 | 104.3 | 97.3-414.1 | 100 | 9.79 | 2.4 | 7.2-12.1 | 50 |
| | 5-OH-Diclofenac | 739.78 | 505.6 | 387.9-1484.9 | 40 | - | - | - | - |
| | Diclofenac–glucuronide | - | - | - | - | 1.41 | - | 1.41 | 25 |
| | Diclofenac-amide | - | - | - | - | - | - | - | - |
| | Acetaminophen | 2035.94 | 3164.1 | 351.2-6779.1 | 40 | 136.13 | 131.0 | 35.7-284.3 | 50 |
| | Acetaminophen–OH | - | - | - | - | 842.71 | 246.2 | 659.3-1269.5 | 100 |
| | Ibuprofen | 4288.07 | 6478.2 | 329.7-21474-1 | 100 | 83.98 | 52.8 | 36.7-171.8 | 100 |
| | 1-OH-ibuprofen | 4707.97 | 3573.3 | 990.9-8117.8 | 30 | 148.63 | 127.8 | 37.6-365.3 | 100 |
| | 2-OH-ibuprofen | 53145.73 | 40840.1 | 15889.4-147755.1 | 100 | 36.97 | 26.9 | 5.7-66.6 | 75 |
| | Carboxy-ibuprofen | 25215.22 | 10481.5 | 18154-40779 | 40 | 137.65 | 145.3 | 15.1-341.6 | 100 |
| β-blocking agents | Metoprolol | 81.71 | - | - | 10 | - | - | - | - |
| | Metoprolol–acid | 118.76 | 71.9 | 16.6-223.4 | 100 | 69.04 | 22.3 | 51.4-107.5 | 100 |
| | Alfa–Metoprolol–OH | - | - | - | - | - | - | - | - |
| | O–Desmethyl–Metoprolol | - | - | - | - | - | - | - | - |

**3.3. Removal of target PhACs**

Removal of anthropogenic contaminants in biological treatment systems comprise mainly biodegradation, bioadsorption and/or bioaccumulation and, in the case of HRAPs, photodegradation should not be neglected either. Considering the operational HRT of 4 d in both ponds, removal rates were calculated using the average influent concentration of 4 consecutive days, and the effluent of the fifth day, as shown in [4]:

$$Removal\ (\%) = 100\ x\ \frac{(\bar{C}_{inf} - C_{eff})}{\bar{C}_{inf}} \qquad [4]$$

where $C_{inf}$ is the average influent concentration of four consecutive days and $C_{eff}$ the effluent concentration of the fifth day. A total of 6 removal values (for each HRAP) were estimated during the two weeks of sampling. In order to get representative results, removals were only estimated for those compounds detected in the influent and effluent wastewater at least 4 days out of the 10 days of sampling. A comparison between the removal efficiencies of HRAPs with and without primary treatment was also carried out, performing a *t*-student analytical test (values are given in SI).

The removals in both HRAP systems are depicted in Figure 4. The elimination of the antibiotic SMX and its metabolite AcSMX were higher in HRAP-test (average values of 85.3% and 89.2%, respectively) than in HRAP-control (50.5% and 82.4%), although in both cases, differences were not significant ($p>0.05$). Due to the higher TSS concentration in the pond without primary treatment, it could be assumed that direct photodegradation did not play a relevant role in the elimination of this antibiotic. However, microalgae can become naturally occurring photosensitizers, generating oxidant species such as hydroxyl radicals and singlet oxygen leading to indirect photolysis of different PhACs (Fatta-Kassinos et al., 2011; Ge et al., 2009). Indeed, Bai and Acharya (2017) observed that the organic matter associated with the microalgal cells could induce indirect photolysis of SMX. Likewise,

SMX is highly water soluble and, at the pond's pH, it is in its anionic form (more stable), which could justify a limited bioadsorption/bioaccumulation to algal cells due to electrostatic repulsion (algal cells are also negatively charged). Concentrations found in the biomass are low, with barely 30 ng $g^{-1}$ and 50 ng $g^{-1}$ in HRAP-control and HRAP-test biomass, respectively. These results suggest that biodegradation/bioassimilation should be considered as the main elimination route of this antibiotic during microalgae treatment, neglecting the possibility of sorption to biomass. Indeed, previous studies demonstrated the low efficiency of green algae (*Nannochloris sp.*) to remove SMX (30%) by bioadsorption under lab-conditions (Bai and Acharya, 2016). Its removal during CAS treatment is quite variable, usually ranging between 50% and 70%, showing also low $K_d$ values on sewage sludge (García-Galán et al., 2011; Trinh et al., 2016). Advanced treatment techniques such as membrane bioreactors (MBRs) do not seem to improve these efficiencies, yielding similar eliminations (García Galán et al., 2012; Mamo et al., 2018). Only the combined use of MBRs followed by reverse osmosis or nanofiltration has proved to be more efficient, achieving removals near 100% (Dolar et al., 2012; Mamo et al., 2018). On the contrary, ozonation, chlorination or indirect photodegradation are highly effective in removing this antibiotic, but pose the threat of the potential toxicity of the TPs generated (Baran et al., 2006; Zessel et al., 2014).

The elimination of the antibiotic metronidazole (MTZ) was high in both HRAPs (>89% in HRAP-control and >91% in HRAP-test), and no significant difference was observed between both ponds. These results are in accordance with previous results in HRAPs with a pretreatment configuration (Díaz-Garduño et al., 2018). However, MTZ was not detected in any biomass sample, being sorption discarded as elimination mechanism. The removal of MTZ during CAS treatment is generally incomplete, with R% usually < 30% as demonstrated in different studies (Dolar et al., 2012; Gros et al., 2012; Jelic et al., 2011).

This negligible elimination agrees with its resilience to biodegradation observed in standardized close bottle test (OECD 301D) (Alexy et al., 2004). On the contrary, the photosensitivity of this antibiotic has been previously demonstrated (Tong et al., 2011), and it is considered to be fully mineralized once discharged on natural waters. Therefore, photodegradation seems to be the most likely removal mechanism in the HRAPs (Figure 5).

Regarding β-blocking agents, the removal of MTP in HRAP-control was very low and highly variable, ranging from no elimination to a maximum of 36%. MTPA was not fully removed in any of the ponds and it showed also a high variability, with values in HRAP-control ranging from no removal to 100%. In HRAP-test, however, only one value out of six was positive (32.3%) for the metabolite. Higher concentrations of MTPA in the treatment effluent cannot be attributed to MTP biodegradation/biotransformation, as it was barely removed and/or not present in the influent (in the case of HRAP-test). Furthermore, MTPA was detected in all the biomass samples, at concentrations between 30-65 ng $g^{-1}$. Previous studies carried out in WWTPs justified these higher concentrations of MTPA in effluent wastewaters to primary degradation of atenolol and not only MTP, as MTPA is a common biodegradation product of both β-blocking agents (Mamo et al., 2018; Radjenović et al., 2008; Rubirola et al., 2014). Despite it was out of the scope of this study, atenolol is more frequently detected than MTP in urban wastewaters (Dolar et al., 2012; Gros et al., 2013) and the presence of MTPA in effluent wastewaters could be attributed to the background presence and degradation of atenolol during microalgae treatment (Díaz-Garduño et al., 2018; Villar-Navarro et al., 2018). Up to date, only fungal treatment has seemed effective in the removal of MTPA, but under lab-scale conditions (Jaén-Gil et al., 2019).

VFX showed positive but limited overall removals in both ponds, with a better elimination in HRAP-control (49%-67%) than in HRAP-test (17%-53%), although differences were not significant in average (p>0.05). Its main metabolites, O-desVFX and N-desVFX, also showed low removal rates (13-39% and 27-42%, respectively), but a much higher variability. In contrast, previous studies on the elimination of these compounds during CAS or MBR treatments demonstrated their limited sorption to sludge (low $K_d$) and their resilience to biodegradation (García-Galán et al., 2016; Mamo et al., 2018). Photodegradation could be considered as the main elimination route for VFX in these systems, and of higher relevance in HRAP-control than in HRAP-test, due to the lower TSS and higher light penetration. The high photodegradability in natural waters of VFX and O-desVFX has already been demonstrated by Rúa-Gómez and Püttmann (2013), who even concluded that once discharged in natural waters these compounds would not pose an environmental risk due to this removal pathway, despite its low biodegradability. On the other hand, the levels found in the biomass for these three compounds were the highest of all the compounds evaluated (see Figure 5). As explained by Santos et al., (2019), and contrary to SMX, VFX is positively charged at pH 8.1, and a strong sorption may occur due to electrostatic forces with the biomass (negatively charged). Regarding the two minor metabolites of VFX detected in the systems, N,N-ddVFX was almost completely removed in both HRAP-control and HRAP-test (average R% of 85% and 88.6%, respectively) whereas N,O-ddVFX was only removed in HRAP-control (65%) and remained in HRAP-test. This metabolite is probably a secondary product of the degradation of the main metabolites of VFX, leading to its presence in effluent wastewaters. The removal of N,O-ddVFX was significantly different between both ponds ($p<0.05$).

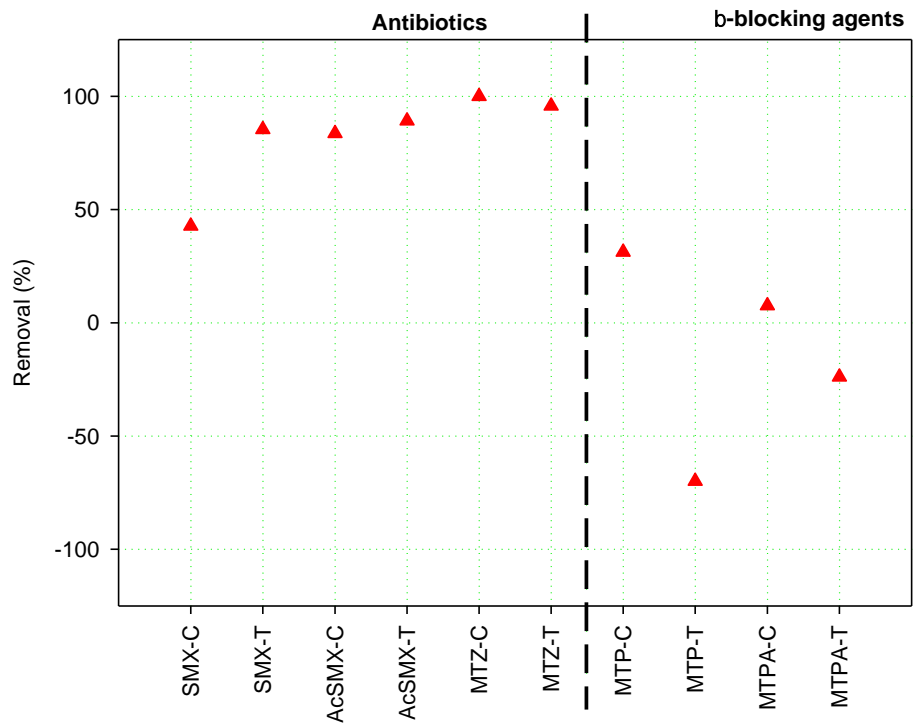

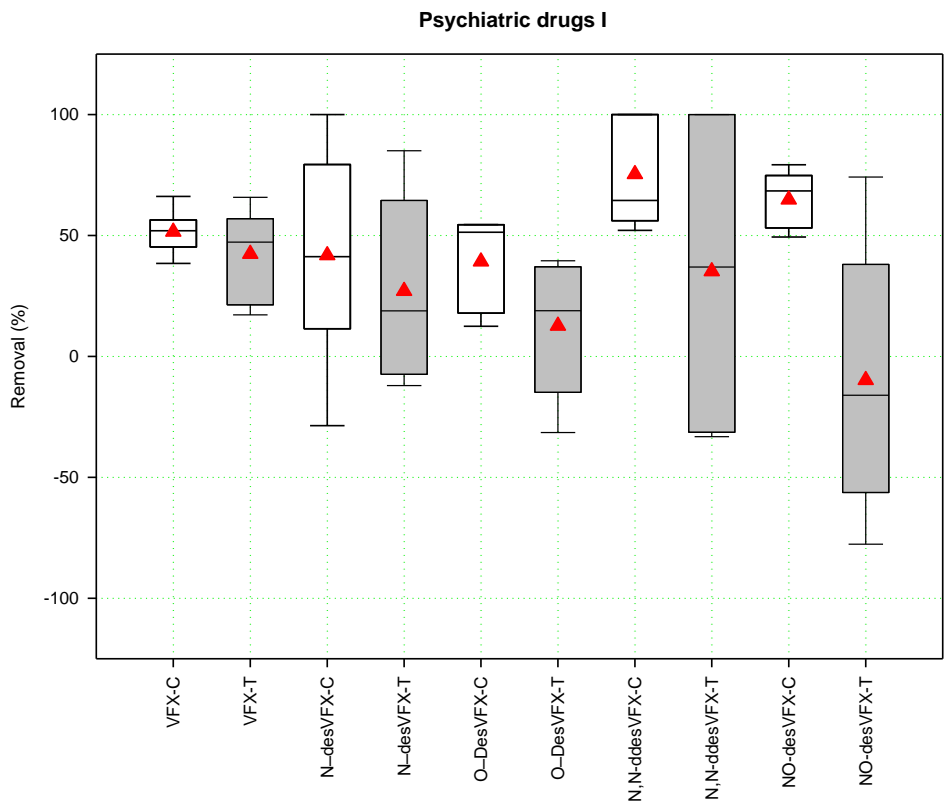

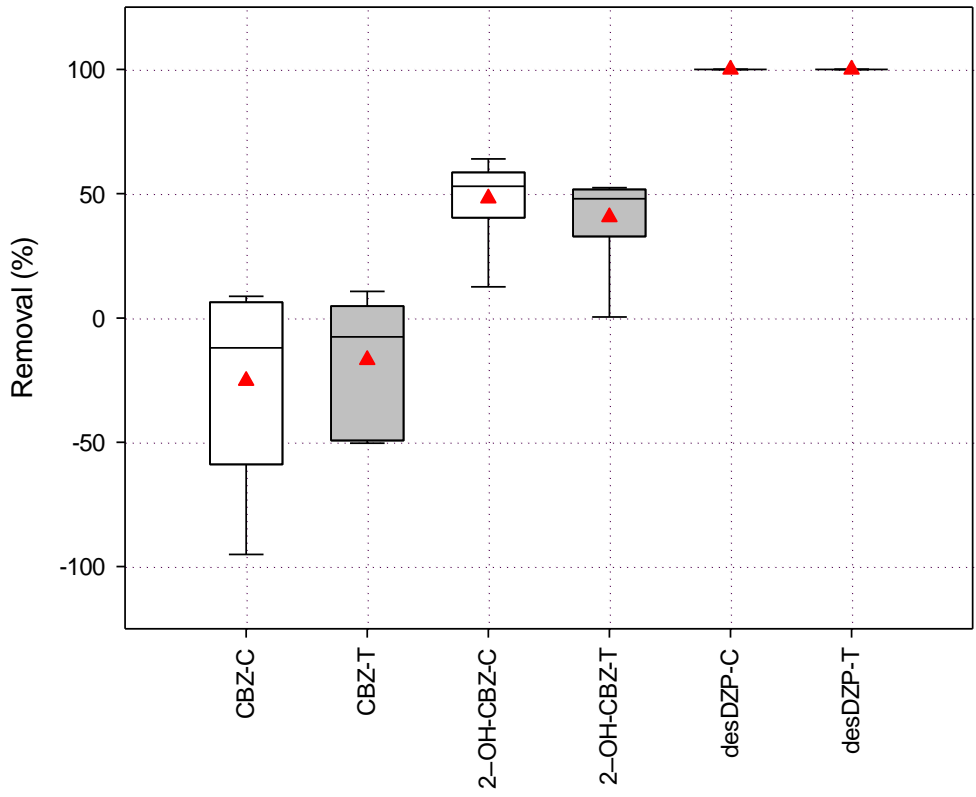

Psychiatric drugs II

Analgesics/anti-inflammatories

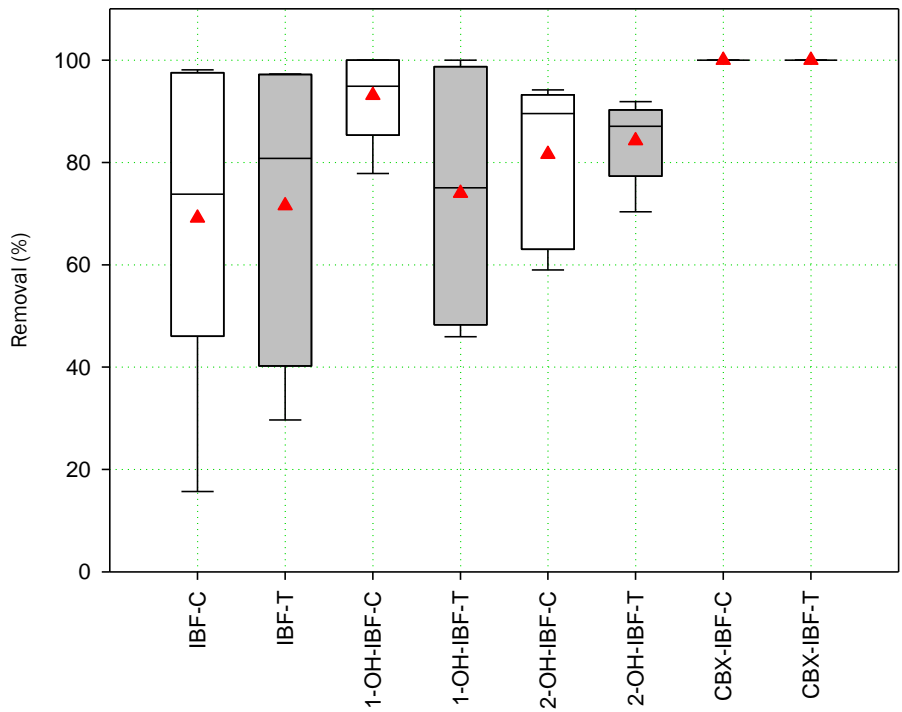

Analgesics/anti-inflammatories II

**Figure 4.** Removals (R%) obtained for the different families of pharmaceuticals in HRAP with primary treatment (white boxes) and HRAP without primary treatment (grey boxes). ▲: average value (the horizontal line is the median value). C and T at the end of the name of the compounds refer to HRAP-control and HRAP-test, respectively.

Regarding CBZ, no removal was observed in either of the ponds. Different studies have reported low removals of CBZ in HRAPs with a pretreatment configuration, ranging from 9% to 23% with HRT of 6 d (Díaz-Garduño et al., 2018; Villar-Navarro et al., 2018). Matamoros et al., (2015) obtained removals of 46% (4 d of HRT) and 62% (8 d HRT) during the warm season, highlighting that even under the best conditions for HRAP efficiency (summer campaigns), CBZ seems to remain unaffected. CBZ elimination in CAS treatments is also negligible, as demonstrated in several studies (Hai et al., 2018). It should be considered that gluruconide moieties of CBZ have never been included in monitoring studies due to the lack of commercial standards. Deconjugation of these metabolites into the original parent compound could explain the regular presence of CBZ in effluents of the different wastewater treatments systems, including HRAPs. Indeed, Vieno et al., (2007) demonstrated the cleavage of CBZ gluruconide forms during CAS wastewater treatment, and their decrease in concentration. On the contrary, the removal of its hydroxylated metabolite was positive in both cases, with values >50% in HRAP-control and between 38-54% in HRAP-test. On the other hand, CBZ was detected in all the biomass samples, at slightly higher concentrations in HRAP-control (19-36.5 ng $g_{-1}$) than in HRAP-test (13-18 ng $g_{-1}$). Acridone, a photodegradation byproduct of CBZ, was also present in all the biomass samples despite not being detected in the influent wastewater to the ponds, and the negligible removal of CBZ. In recent studies carried out under lab-controlled conditions with green algae *Chlorella* sp. and *Scenedesmus* sp., a 20% removal of CBZ was achieved (Matamoros et al., 2016) , and Xiong et al., (2016) obtained CBZ removals up to 37% by freshwater microalgae *Chlamydomonas mexicana* and *Scenedesmus obliquus*. In this last study, the authors concluded that bioadsorption and bioaccumulation were negligible, being biodegradation the main elimination mechanism. De Wilt et al., (2016) obtained eliminations between 10% and 30% in different lab-scale batch experiments with green microalgae *Chlorella sorokiniana*, reaching similar conclusions. Low removals of CBZ are also frequently reported after CAS treatment, and it is considered as a non-degradable, recalcitrant compound (Hai et al., 2018). All in all, the results obtained in this study are consistent with these previous studies and again demonstrate the stability of this compound towards photodegradation and aerobic biodegradation (both in CAS and microalgae treatments). CBZ represents a real challenge in achieving an efficient elimination strategy in real scenarios, and the role of the metabolites in the elimination rate of the parent compound should be considered.

ACM was completely removed in both ponds and not detected in biomass. On the contrary, its main metabolite 3-OH-acetaminophen (ACM-OH) was only detected in the biomass of both ponds, reaching concentrations >800 ng $g_{-1}$ in the HRAP-test. Previous studies also demonstrated a rapid elimination of ACM in HRAPs with primary treatment (Matamoros et al., 2015; Villar-Navarro et al., 2018) and also by *Chlorella sorokiniana* in different batch experiments at lab scale (de Wilt et al., 2016). ACM is a readily biodegradable compound which is fully eliminated after CAS wastewater treatments. Furthermore, it has been concluded that it is photodegradable in surface waters, so that both mechanisms should be pointed out in HRAPs processes. Regarding AcSAc, it showed a very high elimination variability in HRAP-control, with much higher concentrations in the effluents of the pond. This compound was not eliminated either in previous studies with HRAPs (Díaz-Garduño et al., 2018). The presence of AcSAc in biomass was also negligible.

DCF showed an incomplete removal in both ponds, averaging 54.8% in HRAP-control and 51.3% in HRAP-test. It was detected in biomass samples at quite high concentrations too, especially in HRAP-test, so that bioadsorption/bioaccumulation could account partly for its elimination. However, de Wilt et al., (2016) obtained similar removals for DCF (40-60%) in batch reactors using *Chlorella sorokiniana* and attributed its elimination to phototransformation, as it was removed in batches without microalgae inoculum. Indeed, photodegradation of DCF in surface waters has been previously reported (Kunkel and Radke, 2012; Zhang et al., 2008) and also in HRAPs, where DCF removal was considerably higher during the warm/summer season (Matamoros et al., 2015). Considering the fate of both hydroxylated metabolites, 4-OH-DCF and 5-OH-DCF, these were almost completely removed in both ponds, and only 4-OH-DCF was detected in biomass from HRAP-test at levels < 12 ng $g^{-1}$. The glucuronide metabolite of DCF was included in the scope of this study, but not detected in any of the influent samples. All in all, the removal rates observed are similar to those typically obtained in CAS or MBR treatments, due to its generally low biodegradability.

Average removal rates for IBF were 79% and 78% for HRAP-control and HRAP-test, respectively. The removal of the metabolites was also high in both ponds, and the differences were not significant ($p>0.05$) except for cbx-IBF. Ding et al., (2017) obtained lower removals for IBF using the fresh-water diatom *Navicula* sp. under lab-controlled conditions (R% ranging from 20% to 60%); however, these authors worked at a much higher initial concentration of IBF than the environmental levels reported in the present study. Surprisingly, they also observed that adding *Navicula* sp. to the reactors decreased IBF degradation and extended its persistence in the water media. In a different study, IBF removal in the presence of microalgae was also attributed to indirect photodegradation rather than to sorption, due to the presence of dissolved organic matter acting as photocatalysts of the reaction (de Wilt et al., 2016). In fact, the concentration in the biomass of IBF in both HRAPs was not high (Table 3), and higher concentrations were found for its metabolites. The high biodegradability of IBF, which is usually efficiently eliminated during CAS treatment (>90%) explains this low accumulation rate in the biomass (Ferrando-Climent et al., 2012). On the contrary, high removals were obtained for 1-OH IBF in both HRAPs, whereas it exhibited a low elimination in CAS.

Eventually, the removal data gathered from both HRAPs was used to perform a mass balance estimation (Figure 5), including the concentrations found in the algal biomass. The biomass concentration (ng $g^{-1}$) was multiplied by TSS concentration (g $L^{-1}$), in order to homogenize the units of water and biomass concentration in the whole pond system. Therefore, in the present study only the elimination via adsorption or absorption to the microalgae biomass could be quantified. Elimination via photodegradation or biodegradation could not be calculated individually for each target PhAC, but we considered these removal pathways referencing previous research studies as explained previously in this section. For some of the compounds, such as the metabolites of VFX, concentrations in the effluent of the HRAPs were usually higher than that of the influent, and the concentration found in the biomass too. For instance, the concentration of N-desVFX in the HRAP-test effluent was more than 2 times higher than that of the influent, and 4 times higher in the biomass. Degradation of VFX and some of its metabolites by microalgae may lead to the formation of N-desVFX, explaining the high bioaccumulation observed and null removal. In the case of CBZ, effluent values were even 4 times higher (despite influent concentrations of the ponds were very low, < 50 ng $L^{-1}$).

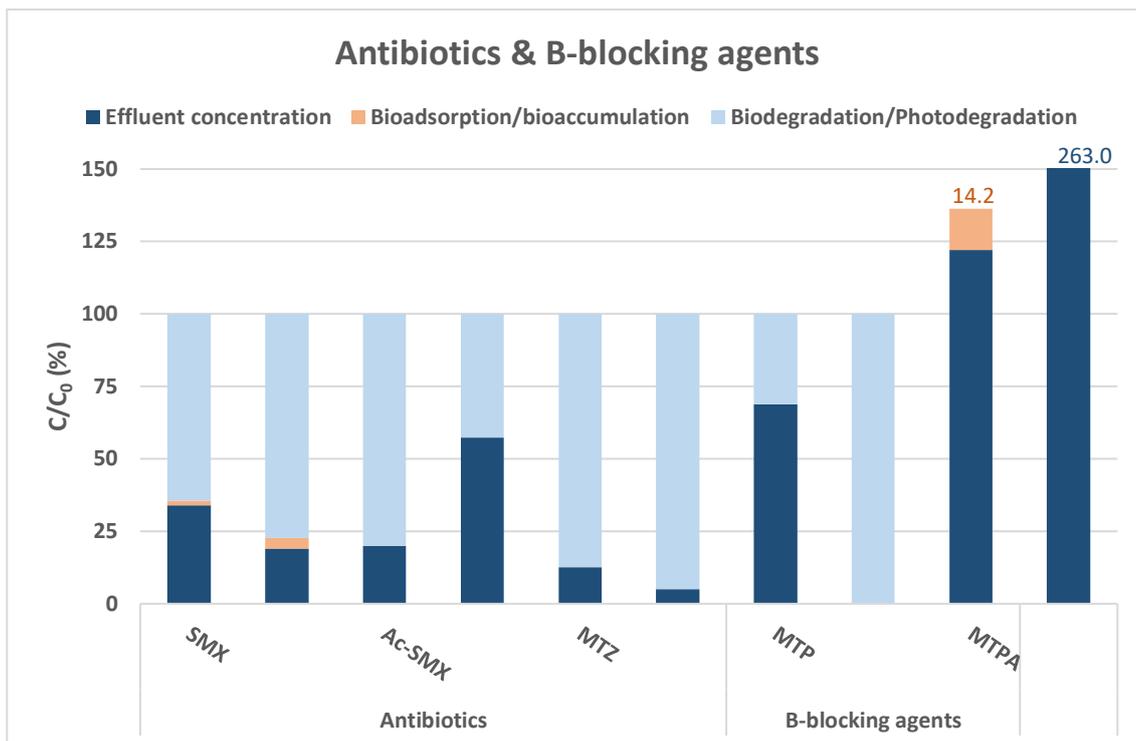

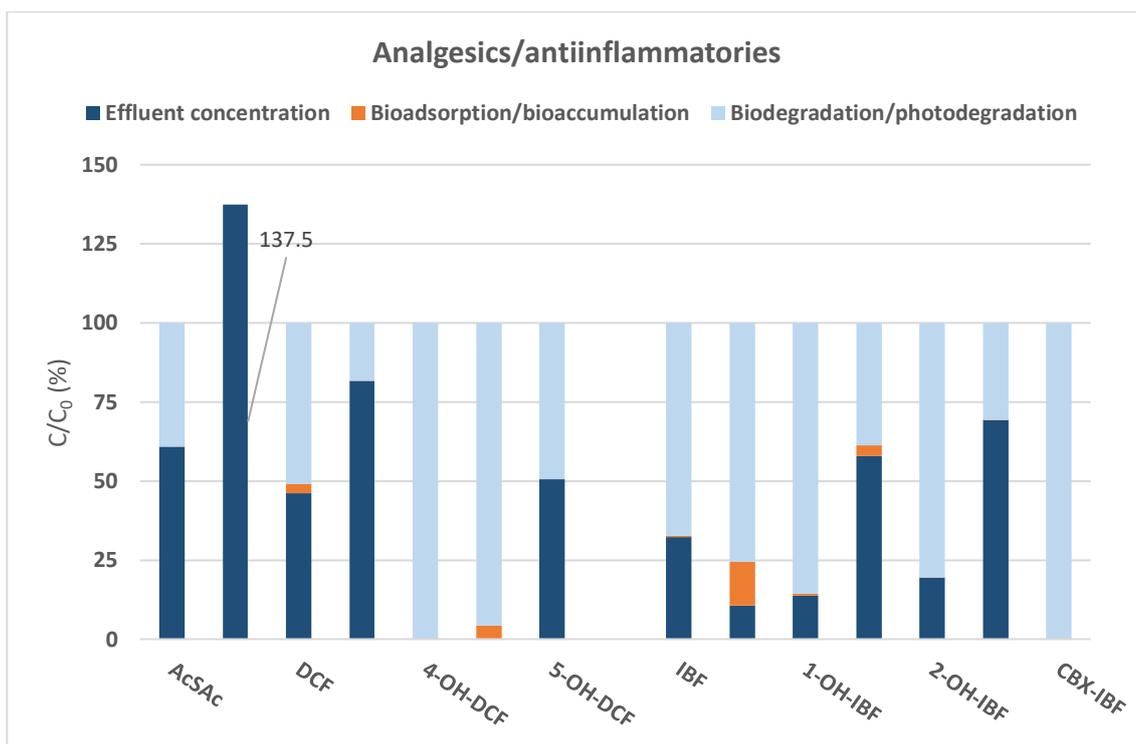

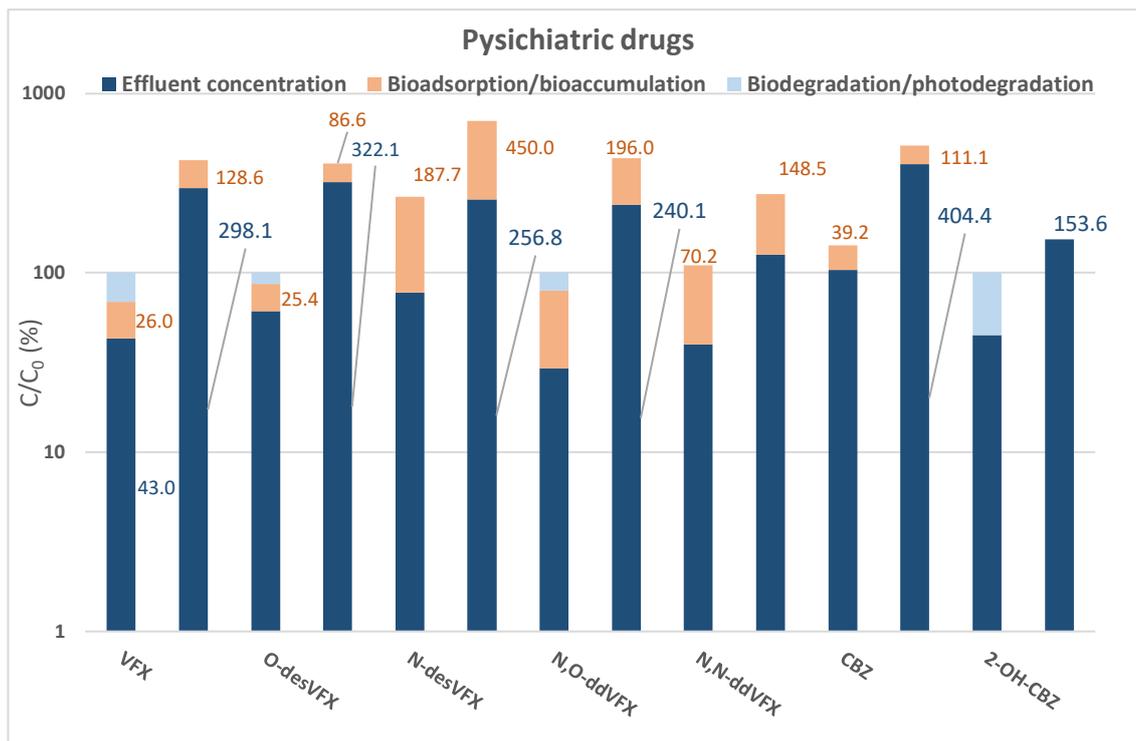

**Figure 5.** Mass balance of the evaluated pharmaceuticals and main metabolites in both HRAPs (given as % of the influent concentration). For each compound, the first data column corresponds to the HRAP HRAP-control and the second to HRAP-test. The balance was performed with the results obtained the 4 days of biomass sampling (so some of the average removal values may differ from those showed in Figure 4).

### 3.4. Biomass

To better understand the bioaccumulation potential of the targeted compounds in the algal biomass, bioaccumulation factors (BAF) were calculated using equation [5]:

$$BAF\ (L\ g^{-1}\ ww) = \frac{C_{biomass}}{C_{water}} \quad [5]$$

where $C_{biomass}$ is the concentrations found in the biomass (ng $g^{-1}$ (wet weight)) and $C_{water}$ the concentration in the effluent wastewater of the pond. As biomass concentration was analyzed 4 days out of the 10 days of sampling, 4 BAFs were calculated for each compound, and are shown in Table 4.

**Table 4.** Bioaccumulation factors (BAF) calculated for the studied compounds in the microalgal biomass (n=4). BAFs are expressed as L kg$_{-1}$ of wet weight (ww) and dry weight (dw). Only values estimated more than two times are included.

| | | HRAP-control | | | | HRAP-test | | | |
|---|---|---|---|---|---|---|---|---|---|
| | Compound | BAF$_{ww}$ | SD | BAF$_{dw}$ | SD | BAF$_{ww}$ | SD | BAF$_{dw}$ | SD |
| Antibiotics | Sulfamethoxazole | 144,05 | 9,0 | 1642,16 | 102,7 | 35,70 | 0,4 | 407,17 | 5,0 |
| Psychiatric drugs | Venlafaxine | 170,7 | 19,3 | 1945,8 | 220,5 | 80,3 | 23,0 | 915,22 | 262,6 |
| | N–Desmethylvenlafaxine | 419,63 | 5,0 | 4783,7 | 57,4 | 287,6 | 66,1 | 3279,11 | 753,2 |
| | O–Desmethylvenlafaxine | 100,7 | 25,0 | 1147,8 | 285,2 | 52,7 | 8,4 | 600,50 | 96,3 |
| | N,N-Didesmethylvenlafaxine | 267,1 | 225,0 | 3044,9 | 2565,4 | 222,5 | 2,9 | 2536,76 | 32,9 |
| | NO-desmethylvenlafaxine | 395,6 | 146,9 | 4510,1 | 1674,4 | 142,5 | 13,7 | 1624,75 | 156,4 |
| | Carbamazepine | 81,3 | 23,9 | 926,3 | 272,9 | 50,1 | 5,7 | 571,21 | 64,8 |
| | 2–Hydroxycarbamazepine | - | - | - | - | 39,8 | 8,6 | 453,26 | 98,3 |
| | Acridone | 914,3 | 192,0 | 10422,8 | 2188,7 | - | - | - | - |
| Analgesics/antiinflammatories | Diclofenac | 14,4 | 19,0 | 164,2 | 216,4 | 54,9 | 11,0 | 626,01 | 125,4 |
| | 4-OH-Diclofenac | - | - | - | - | 17,8 | 2.1 | 203,19 | 23.1 |
| | Ibuprofen | 5,2 | 7,0 | 59,1 | 79,9 | 10,2 | 4,7 | 116,32 | 54,1 |
| | 1-OH-ibuprofen | 18,5 | 7,9 | 210,7 | 90,2 | 13,2 | 4,9 | 150,07 | 56,5 |
| | 2-OH-ibuprofen | 0,3 | 0,4 | 3,7 | 4,3 | 0,08 | 0,01 | 0,96 | 0,1 |
| β-blocking agents | Metoprolol–acid | 30,81 | 10,93 | 351,29 | 124,62 | 41,61 | 15,4 | 474,35 | 175,9 |

The highest BAF values corresponded to acridone, a photodegradation byproduct of CBZ, which was detected in all the biomass samples and barely in effluent wastewaters (values < 10 ng L$_{-1}$). VFX and its metabolites also showed high BAF values, due to their low biodegradability in both HRAPs, as explained in section 3.4. The accumulation potential of this drug has been addressed in previous studies in freshwater biofilm, obtaining BAFs in the range of 700-3000 L kg$_{-1(dw)}$ in a WWTP-impacted river (Huerta et al., 2016). These values are higher than those obtained for microalgae, due to the chronic exposure of this biofilm to the polluted river water. Santos et al., (2019) also obtained results in that range in river biofilm (960 L kg$_{-1(dw)}$) in mesocosm-scale experiments after 72 h of exposure to an initial concentration of 50 µg L$_{-1}$. SMX also showed a high tendency to be sorbed onto the biomass, with higher BAFs in microalgae than that observed in benthic invertebrates after lab-scale exposure experiments to 1 µg L$_{-1}$ (Garcia-Galan et al., 2017). MTPA, which was present at higher concentrations in both HRAPs effluents than in the influent, also showed relatively high BAFs. Bioaccumulation potential of DCF in microalgae ranged from 164 to 626 L kg$_{-1(dw)}$ and was lower than that observed in freshwater biofilm (920-3540 L kg$_{-1}$), probably due to the shorter exposure in the HRAPs. For the rest of the compounds studied, and especially for the analgesics/anti-inflammatories, which are generally highly biodegradable, BAFs were low/negligible. BAFs were generally higher in the HRAP-control than in HRAP-test; despite the slightly higher biomass concentrations in HRAP-test for most of the compounds, the higher variability and lower removal rates in this test pond may be responsible of this lower BAFs. Nevertheless, the differences in the removal efficiency between both HRAPs were not significant. To the authors' knowledge, there are no previous BAFs values of PhACs in microalgae biomass available in the literature.

## 4. CONCLUSION

The present study has demonstrated that the implementation of HRAPs as secondary treatment is a feasible alternative to CAS treatment, in terms of both overall wastewater treatment (COD, NH$_4^+$-N, TN, organic matter) and also of organic micropollutants removal, with generally higher or similar removal performances. HRAPs can be operated without primary treatment, without significant changes in the overall performance and saving costs. These ponds have shown high removal efficiencies for a wide range of PhACs, including many of their metabolites and transformation products. Furthermore, the removal of the primary treatment that generally precedes HRAPs, seemed to have no effect in the performance of the ponds regarding both PhACs elimination and overall wastewater treatment efficiency. HRAPs configuration without a primary settler could lead to savings in system maintenance and costs. However, a higher productivity of biomass in HRAP-control than in HRAP-test was observed during the sampling campaign, with values up to 3 times higher (14.6 gVSS $m^{-2} \cdot d^{-1}$ vs 5 gVSS $m^{-2} \cdot d^{-1}$), due to the light blockage in HRAP-test that reduced biomass growth. One of the main advantages of using microalgae in wastewater treatment is indeed that the biomass produced can be harvested to obtain biofertilizers (as it recycles nutrients such as nitrogen or phosphorus), bioenergy and other added-value products etc, so a higher biomass production is very convenient.

The removal of the target PhACs and the corresponding metabolites in both HRAPs has usually ranged from moderate to high. For some recalcitrant compounds such as CBZ or MTP and its metabolite, MTPA, removal efficiencies did not improved from those obtained in CAS treatment. On the contrary, for some compounds such as VFX, O-desVFX or MTZ, which are barely removed during CAS treatments, removals were higher, resulting in values of 50%, 35% and 100% in average, respectively. Moderate removals, such as those of DCF or SMX, can be explained in terms of back-transformation in the parent drug of their corresponding acetylated or glucuronidated metabolites. On the contrary, for some compounds such as VFX, O-desVFX or MTZ, which are barely removed during CAS treatments, removals were higher, resulting in values of 50%, 35% and 100% in average, respectively. Occurrence of the target analytes in microalgae biomass has also been evaluated, in order to discern if sorption had been the main removal mechanism during treatment. In the case of VFX and O-desVFX, bioadsorption/bioaccumulation in the microalgae biomass may have played a decisive role in their elimination. As expected, the less biodegradable compounds showed the highest tendencies to be adsorbed onto the biomass (highest BAFs). To the author's knowledge, this is the first time that BAF values are estimated for microalgae biomass.

All in all, data on PhACs removal in HRAPs under real conditions (environmental concentrations) is still scarce, as most of the studies carried out are lab-scaled, with pure cultures, sterile conditions and at much higher doped concentrations of the contaminant. To the author's knowledge, this is the first study dealing with the occurrence of PhACs and metabolites in microalgae biomass under environmental conditions.

**Acknowledgements**

This research was funded by the Spanish Ministry of Economy and Competitiveness (FOTOBIOGAS Project CTQ2014-57293-C3-3-R) and the European Union's Horizon 2020 research and innovation program under the Marie Skłodowska-Curie grant agreement No 676070 (SuPER-W). M.J. García-Galán and L.H.M.L.M. Santos acknowledges their Juan de la Cierva research grants (FJCI-2014-22767, IJCI-2017-34601 and IJCI-2017-32747, respectively), and M. Garfí and S. Rodríguez-Mozaz thanks the Ramon y Cajal program (RYC-2016-20059 and RYC-2014-16707, respectively), all from the Spanish Ministry of Economy and Competitiveness.

**REFERENCES**

Abdel-Raouf, N., Al-Homaidan, A.A., Ibraheem, I.B.M., 2012. Microalgae and wastewater treatment. Saudi J. Biol. Sci. 19, 257–275. https://doi.org/http://dx.doi.org/10.1016/j.sjbs.2012.04.005

AEMPS, 2012. www.aemps.gob.es/cima/especialidad.do?metodo=verPresentaciones&codigo=75561 [WWW Document]. ぎょうせい.

Alexy, R., Kümpel, T., Kümmerer, K., 2004. Assessment of degradation of 18 antibiotics in the Closed Bottle Test. Chemosphere. https://doi.org/10.1016/j.chemosphere.2004.06.024

APHA-AWWA-WEF, 2012. Standard methods for the examination of water and wastewater, 22nd ed. Published jointly by the American Water Works Association, the American Public Health Association, and the Water Environment Federation., Washington, D.C.

Arashiro, L.T., Ferrer, I., Rousseau, D.P.L., Van Hulle, S.W.H., Garfí, M., 2019. The effect of primary treatment of wastewater in high rate algal pond systems: biomass and bioenergy recovery. Bioresour. Technol. 280, 27–36. https://doi.org/https://doi.org/10.1016/j.biortech.2019.01.096

Archana, G., Dhodapkar, R., Kumar, A., 2017. Ecotoxicological risk assessment and seasonal variation of some pharmaceuticals and personal care products in the sewage treatment plant and surface water bodies (lakes). Environ. Monit. Assess. 189. https://doi.org/10.1007/s10661-017-6148-3

Arias, D.M., Uggetti, E., García-Galán, M.J., García, J., 2018. Production of polyhydroxybutyrates and carbohydrates in a mixed cyanobacterial culture: Effect of nutrients limitation and photoperiods. N. Biotechnol. 42, 1–11. https://doi.org/10.1016/j.nbt.2018.01.001

Aristi, I., von Schiller, D., Arroita, M., Barceló, D., Ponsatí, L., García-Galán, M.J., Sabater, S., Elosegi, A., Acuña, V., 2015. Mixed effects of effluents from a wastewater treatment plant on river ecosystem metabolism: Subsidy or stress? Freshw. Biol. 60. https://doi.org/10.1111/fwb.12576

Bai, X., Acharya, K., 2017. Algae-mediated removal of selected pharmaceutical and personal care products (PPCPs) from Lake Mead water. Sci. Total Environ. https://doi.org/10.1016/j.scitotenv.2016.12.192

Bai, X., Acharya, K., 2016. Removal of trimethoprim, sulfamethoxazole, and triclosan by the green alga Nannochloris sp. J. Hazard. Mater. https://doi.org/10.1016/j.jhazmat.2016.04.067

Baran, W., Sochacka, J., Wardas, W., 2006. Toxicity and biodegradability of sulfonamides and products of their photocatalytic degradation in aqueous solutions. Chemosphere. https://doi.org/10.1016/j.chemosphere.2006.04.040

Benemann, J.R., Koopman, B., Weissman, J., Oswald, W.J., 2013. BIOMASS PRODUCTION AND WASTE RECYCLING WITH BLUE-GREEN ALGAE, in: Microbial Energy Conversion. https://doi.org/10.1016/b978-0-08-021791-8.50037-8

Bourrelly, P., 1990. Les algues d'eau douce. Les Algues Vertes.

Bradley, P.M., Battaglin, W.A., Clark, J.M., Henning, F.P., Hladik, M.L., Iwanowicz, L.R., Journey, C.A., Riley, J.W., Romanok, K.M., 2017. Widespread occurrence and potential for biodegradation of bioactive contaminants in Congaree National Park, USA. Environ. Toxicol. Chem. 36, 3045–3056. https://doi.org/10.1002/etc.3873

Chisti, Y., 2013. Raceways-based production of algal crude oil. Green 3, 195–216. https://doi.org/10.1515/green-2013-0018

de Godos, I., Arbib, Z., Lara, E., Rogalla, F., 2016. Evaluation of High Rate Algae Ponds for treatment of anaerobically digested wastewater: Effect of CO2 addition and modification of dilution rate. Bioresour. Technol. 220, 253–261. https://doi.org/10.1016/j.biortech.2016.08.056

de Wilt, A., Butkovskyi, A., Tuantet, K., Leal, L.H., Fernandes, T. V., Langenhoff, A., Zeeman, G., 2016. Micropollutant removal in an algal treatment system fed with source separated wastewater streams. J. Hazard. Mater. https://doi.org/10.1016/j.jhazmat.2015.10.033

Díaz-Garduño, B., Perales, J.A., Biel-Maeso, M., Pintado-Herrera, M.G., Lara-Martin, P.A.,

Garrido-Pérez, C., Martín-Díaz, M.L., 2018. Biochemical responses of Solea senegalensis after continuous flow exposure to urban effluents. Sci. Total Environ. https://doi.org/10.1016/j.scitotenv.2017.09.304

Ding, T., Yang, M., Zhang, J., Yang, B., Lin, K., Li, J., Gan, J., 2017. Toxicity, degradation and metabolic fate of ibuprofen on freshwater diatom Navicula sp. J. Hazard. Mater. https://doi.org/10.1016/j.jhazmat.2017.02.004

Dolar, D., Gros, M., Rodriguez-Mozaz, S., Moreno, J., Comas, J., Rodriguez-Roda, I., Barceló, D., 2012. Removal of emerging contaminants from municipal wastewater with an integrated membrane system, MBR-RO. J. Hazard. Mater. 239–240, 64–69. https://doi.org/10.1016/j.jhazmat.2012.03.029

Dvořáková Březinova, T., Vymazal, J., Koželuh, M., Kule, L., 2018. Occurrence and removal of ibuprofen and its metabolites in full-scale constructed wetlands treating municipal wastewater. Ecol. Eng. https://doi.org/10.1016/j.ecoleng.2018.05.020

Fatta-Kassinos, D., Vasquez, M.I., Kümmerer, K., 2011. Transformation products of pharmaceuticals in surface waters and wastewater formed during photolysis and advanced oxidation processes - Degradation, elucidation of byproducts and assessment of their biological potency. Chemosphere 85, 693–709.

Ferrando-Climent, L., Collado, N., Buttiglieri, G., Gros, M., Rodriguez-Roda, I., Rodriguez-Mozaz, S., Barceló, D., 2012. Comprehensive study of ibuprofen and its metabolites in activated sludge batch experiments and aquatic environment. Sci. Total Environ. https://doi.org/10.1016/j.scitotenv.2012.08.073

Fisher, D.K., Pringle III, H.C., 2013. Evaluation of alternative methods for estimating reference evapotranspiration. Agric. Sci. 04, 51–60. https://doi.org/10.4236/as.2013.48A008

García-Galán, M.J., Anfruns, A., Gonzalez-Olmos, R., Rodríguez-Mozaz, S., Comas, J., 2016.

UV/H2O2degradation of the antidepressants venlafaxine and O-desmethylvenlafaxine: Elucidation of their transformation pathway and environmental fate. J. Hazard. Mater. 311, 70–80. https://doi.org/http://dx.doi.org/10.1016/j.jhazmat.2016.02.070

García-Galán, M.J., Díaz-Cruz, M.S., Barceló, D., 2011. Occurrence of sulfonamide residues along the Ebro river basin. Removal in wastewater treatment plants and environmental impact assessment. Environ. Int. 37, 462–473. https://doi.org/10.1016/j.envint.2010.11.011

García-Galán, M.J., Gutiérrez, R., Uggetti, E., Matamoros, V., García, J., Ferrer, I., 2018. Use of full-scale hybrid horizontal tubular photobioreactors to process agricultural runoff. Biosyst. Eng. 166, 138–149. https://doi.org/10.1016/j.biosystemseng.2017.11.016

García-Galán, M.J., Petrovic, M., Rodríguez-Mozaz, S., Barceló, D., 2016. Multiresidue trace analysis of pharmaceuticals, their human metabolites and transformation products by fully automated on-line solid-phase extraction-liquid chromatography-tandem mass spectrometry. Talanta 158, 330–341. https://doi.org/10.1016/j.talanta.2016.05.061

García-Galán, M.J., Petrovic, M., Rodríguez-Mozaz, S., Barceló, D., 2016. Multiresidue trace analysis of pharmaceuticals, their human metabolites and transformation products by fully automated on-line solid-phase extraction-liquid chromatography-tandem mass spectrometry. Talanta 158. https://doi.org/10.1016/j.talanta.2016.05.061

Garcia-Galan, Maria Jesus, Sordet, M., Buleté, A., Garric, J., Vulliet, E., 2017. Evaluation of the influence of surfactants in the bioaccumulation kinetics of sulfamethoxazole and oxazepam in benthic invertebrates. Sci. Total Environ. 592, 554–564. https://doi.org/https://doi.org/10.1016/j.scitotenv.2017.03.085

Garcia-Galan, M.J., Sordet, M., Buleté, A., Garric, J., Vulliet, E., 2017. Evaluation of the influence of surfactants in the bioaccumulation kinetics of sulfamethoxazole and oxazepam in benthic invertebrates. Sci. Total Environ. 592.

https://doi.org/10.1016/j.scitotenv.2017.03.085

García Galán, M.J., Díaz-Cruz, M.S., Barceló, D., 2012. Removal of sulfonamide antibiotics upon conventional activated sludge and advanced membrane bioreactor treatment. Anal. Bioanal. Chem. 404. https://doi.org/10.1007/s00216-012-6239-5

García, J., Green, B.F., Lundquist, T., Mujeriego, R., Hernández-Mariné, M., Oswald, W.J., 2006. Long term diurnal variations in contaminant removal in high rate ponds treating urban wastewater. Bioresour. Technol. 97, 1709–1715. https://doi.org/10.1016/j.biortech.2005.07.019

Ge, L., Deng, H., Wu, F., Deng, N., 2009. Microalgae-promoted photodegradation of two endocrine disrupters in aqueous solutions. J. Chem. Technol. Biotechnol. https://doi.org/10.1002/jctb.2043

Gracia-Lor, E., Sancho, J. V., Serrano, R., Hernández, F., 2012. Occurrence and removal of pharmaceuticals in wastewater treatment plants at the Spanish Mediterranean area of Valencia. Chemosphere. https://doi.org/10.1016/j.chemosphere.2011.12.025

Gros, M., Petrović, M., Barceló, D., 2007. Wastewater treatment plants as a pathway for aquatic contamination by pharmaceuticals in the ebro river basin (northeast Spain). Environ. Toxicol. Chem. 26, 1553–1562.

Gros, M., Petrovic, M., Ginebreda, A., Barceló, D., 2010. Removal of pharmaceuticals during wastewater treatment and environmental risk assessment using hazard indexes. Environ. Int. 36, 15–26.

Gros, M., Rodríguez-Mozaz, S., Barceló, D., 2013. Rapid analysis of multiclass antibiotic residues and some of their metabolites in hospital, urban wastewater and river water by ultra-high-performance liquid chromatography coupled to quadrupole-linear ion trap tandem mass spectrometry. J. Chromatogr. A 1292, 173–188. https://doi.org/10.1016/j.chroma.2012.12.072

Gros, M., Rodríguez-Mozaz, S., Barceló, D., 2012. Fast and comprehensive multi-residue analysis of a broad range of human and veterinary pharmaceuticals and some of their metabolites in surface and treated waters by ultra-high-performance liquid chromatography coupled to quadrupole-linear ion trap tandem. J. Chromatogr. A 1248, 104–121. https://doi.org/10.1016/j.chroma.2012.05.084

Guo, X., Pang, W., Dou, C., Yin, D., 2017. Sulfamethoxazole and COD increase abundance of sulfonamide resistance genes and change bacterial community structures within sequencing batch reactors. Chemosphere. https://doi.org/10.1016/j.chemosphere.2017.01.134

Hai, F.I., Yang, S., Asif, M.B., Sencadas, V., Shawkat, S., Sanderson-Smith, M., Gorman, J., Xu, Z.Q., Yamamoto, K., 2018. Carbamazepine as a Possible Anthropogenic Marker in Water: Occurrences, Toxicological Effects, Regulations and Removal by Wastewater Treatment Technologies. Water (Switzerland). https://doi.org/10.3390/w10020107

Hübner, U., von Gunten, U., Jekel, M., 2015. Evaluation of the persistence of transformation products from ozonation of trace organic compounds – A critical review. Water Res. 68, 150–170. https://doi.org/http://dx.doi.org/10.1016/j.watres.2014.09.051

Huerta, B., Jakimska, A., Llorca, M., Ruhí, A., Margoutidis, G., Acuña, V., Sabater, S., Rodriguez-Mozaz, S., Barcelò, D., 2015. Development of an extraction and purification method for the determination of multi-class pharmaceuticals and endocrine disruptors in freshwater invertebrates. Talanta 132, 373–381. https://doi.org/http://dx.doi.org/10.1016/j.talanta.2014.09.017

Huerta, B., Rodriguez-Mozaz, S., Nannou, C., Nakis, L., Ruhí, A., Acuña, V., Sabater, S., Barcelo, D., 2016. Determination of a broad spectrum of pharmaceuticals and endocrine disruptors in biofilm from a waste water treatment plant-impacted river. Sci. Total Environ. https://doi.org/10.1016/j.scitotenv.2015.05.049

Jaén-Gil, A., Castellet-Rovira, F., Llorca, M., Villagrasa, M., Sarrà, M., Rodríguez-Mozaz, S., Barceló, D., 2019. Fungal treatment of metoprolol and its recalcitrant metabolite metoprolol acid in hospital wastewater: Biotransformation, sorption and ecotoxicological impact. Water Res. https://doi.org/10.1016/j.watres.2018.12.054

Jaén-Gil, A., Hom-Diaz, A., Llorca, M., Vicent, T., Blánquez, P., Barceló, D., Rodríguez-Mozaz, S., 2018. An automated on-line turbulent flow liquid-chromatography technology coupled to a high resolution mass spectrometer LTQ-Orbitrap for suspect screening of antibiotic transformation products during microalgae wastewater treatment. J. Chromatogr. A. https://doi.org/10.1016/j.chroma.2018.06.027

Jelic, A., Gros, M., Ginebreda, A., Cespedes-Sánchez, R., Ventura, F., Petrovic, M., Barcelo, D., 2011. Occurrence, partition and removal of pharmaceuticals in sewage water and sludge during wastewater treatment. Water Res. https://doi.org/10.1016/j.watres.2010.11.010

Jelic, A., Rodriguez-Mozaz, S., Barceló, D., Gutierrez, O., 2015. Impact of in-sewer transformation on 43 pharmaceuticals in a pressurized sewer under anaerobic conditions. Water Res. 68, 98–108. https://doi.org/10.1016/j.watres.2014.09.033

Khetkorn, W., Incharoensakdi, A., Lindblad, P., Jantaro, S., 2016. Enhancement of poly-3-hydroxybutyrate production in Synechocystis sp. PCC 6803 by overexpression of its native biosynthetic genes. Bioresour. Technol. 214, 761–768. https://doi.org/10.1016/j.biortech.2016.05.014

Khetkorn, W., Rastogi, R.P., Incharoensakdi, A., Lindblad, P., Madamwar, D., Pandey, A., Larroche, C., 2017. Microalgal hydrogen production – A review. Bioresour. Technol. https://doi.org/10.1016/j.biortech.2017.07.085

Kostich, M.S., Batt, A.L., Lazorchak, J.M., 2014. Concentrations of prioritized pharmaceuticals in ef fl uents from 50 large wastewater treatment plants in the US and implications for risk estimation. Environ. Pollut. 184, 354–359.

https://doi.org/10.1016/j.envpol.2013.09.013

Kunkel, U., Radke, M., 2012. Fate of pharmaceuticals in rivers: Deriving a benchmark dataset at favorable attenuation conditions. Water Res. https://doi.org/10.1016/j.watres.2012.07.033

Langdon, K.A., Warne, M.S.T.J., Kookanaz, R.S., 2010. Aquatic hazard assessment for pharmaceuticals, personal care products, and endocrine-disrupting compounds from biosolids-amended land. Integr. Environ. Assess. Manag. 6, 663–676.

Li, Y., Zhang, L., Liu, X., Ding, J., 2019. Ranking and prioritizing pharmaceuticals in the aquatic environment of China. Sci. Total Environ. 658, 333–342. https://doi.org/10.1016/j.scitotenv.2018.12.048

Madikizela, L.M., Tavengwa, N.T., Chimuka, L., 2017. Status of pharmaceuticals in African water bodies: Occurrence, removal and analytical methods. J. Environ. Manage. 193, 211–220. https://doi.org/10.1016/j.jenvman.2017.02.022

Maestrini, S.Y., 1982. Simultaneous uptake of ammonium and nitrate by oyster-pond algae. Mar. Biol. Lett. 3, 143–153.

Mamo, J., García-Galán, M.J., Stefani, M., Rodríguez-Mozaz, S., Barceló, D., Monclús, H., Rodriguez-Roda, I., Comas, J., 2018. Fate of pharmaceuticals and their transformation products in integrated membrane systems for wastewater reclamation. Chem. Eng. J. 331, 450–461. https://doi.org/10.1016/j.cej.2017.08.050

Matamoros, V., Gutiérrez, R., Ferrer, I., García, J., Bayona, J.M., 2015. Capability of microalgae-based wastewater treatment systems to remove emerging organic contaminants: A pilot-scale study. J. Hazard. Mater. 288, 34–42. https://doi.org/10.1016/j.jhazmat.2015.02.002

Matamoros, V., Uggetti, E., García, J., Bayona, J.M., 2016. Assessment of the mechanisms

involved in the removal of emerging contaminants by microalgae from wastewater: A laboratory scale study. J. Hazard. Mater. https://doi.org/10.1016/j.jhazmat.2015.08.050

Muñoz, R., Guieysse, B., 2006. Algal-bacterial processes for the treatment of hazardous contaminants: A review. Water Res. 40, 2799–2815. https://doi.org/10.1016/j.watres.2006.06.011

Norvill, Z.N., Toledo-Cervantes, A., Blanco, S., Shilton, A., Guieysse, B., Muñoz, R., 2017. Photodegradation and sorption govern tetracycline removal during wastewater treatment in algal ponds. Bioresour. Technol. https://doi.org/10.1016/j.biortech.2017.02.011

Nurdogan, Y., Oswald, W.J., 1995a. Enhanced nutrient removal in high-rate ponds. Water Sci. Technol. 31, 33–43. https://doi.org/10.1016/0273-1223(95)00490-E

Nurdogan, Y., Oswald, W.J., 1995b. Enhanced nutrient removal in high-rate ponds. Water Sci. Technol. https://doi.org/10.1016/0273-1223(95)00490-E

Oaks, J.L., Gilbert, M., Virani, M.Z., Watson, R.T., Meteyer, C.U., Rideout, B.A., Shivaprasad, H.L., Ahmed, S., Chaudhry, M.J.I., Arshad, M., Mahmood, S., Ali, A., Khan, A.A., 2004. Diclofenac residues as the cause of vulture population decline in Pakistan. Nature. https://doi.org/10.1038/nature02317

Oliver, R.L., Ganf, G.G., 2002. Freshwater Blooms BT - The Ecology of Cyanobacteria: Their Diversity in Time and Space, in: Whitton, B.A., Potts, M. (Eds.), . Springer Netherlands, Dordrecht, pp. 149–194. https://doi.org/10.1007/0-306-46855-7_6

Ortiz de García, S., Pinto Pinto, G., García Encina, P., Irusta Mata, R., 2013. Consumption and occurrence of pharmaceutical and personal care products in the aquatic environment in Spain. Sci. Total Environ. 444, 451–465. https://doi.org/10.1016/J.SCITOTENV.2012.11.057

Oswald, W.J., 1995. Ponds in the twenty-first century. Water Sci. Technol. 31, 1–8. https://doi.org/10.1016/0273-1223(95)00487-8

Palmer, C.M., 1962. Algas en abastecimientos de agua: manual ilustrado acerca de la identificación, importancia y control de las algas en los abastecimientos de agua. Interamericana.

Radjenović, J., Pérez, S., Petrović, M., Barceló, D., 2008. Identification and structural characterization of biodegradation products of atenolol and glibenclamide by liquid chromatography coupled to hybrid quadrupole time-of-flight and quadrupole ion trap mass spectrometry. J. Chromatogr. A. https://doi.org/10.1016/j.chroma.2008.09.060

Rodriguez-Mozaz, S., Chamorro, S., Marti, E., Huerta, B., Gros, M., Sànchez-Melsió, A., Borrego, C.M., Barceló, D., Balcázar, J.L., 2015. Occurrence of antibiotics and antibiotic resistance genes in hospital and urban wastewaters and their impact on the receiving river. Water Res. https://doi.org/10.1016/j.watres.2014.11.021

Rúa-Gómez, P.C., Püttmann, W., 2013. Degradation of lidocaine, tramadol, venlafaxine and the metabolites O-desmethyltramadol and O-desmethylvenlafaxine in surface waters. Chemosphere. https://doi.org/10.1016/j.chemosphere.2012.10.039

Rubirola, A., Llorca, M., Rodriguez-Mozaz, S., Casas, N., Rodriguez-Roda, I., Barceló, D., Buttiglieri, G., 2014. Characterization of metoprolol biodegradation and its transformation products generated in activated sludge batch experiments and in full scale WWTPs. Water Res. https://doi.org/10.1016/j.watres.2014.05.031

Ruhí, A., Acuña, V., Barceló, D., Huerta, B., Mor, J.-R., Rodríguez-Mozaz, S., Sabater, S., 2016. Bioaccumulation and trophic magnification of pharmaceuticals and endocrine disruptors in a Mediterranean river food web. Sci. Total Environ. 540, 250–259. https://doi.org/http://dx.doi.org/10.1016/j.scitotenv.2015.06.009

Ruiz-Marin, A., Mendoza-Espinosa, L.G., Stephenson, T., 2010. Growth and nutrient removal in free and immobilized green algae in batch and semi-continuous cultures treating real wastewater. Bioresour. Technol. 101, 58–64.

https://doi.org/10.1016/j.biortech.2009.02.076

Santos, L.H.M.L.M., Freixa, A., Insa, S., Acuña, V., Sanchís, J., Farré, M., Sabater, S., Barceló, D., Rodríguez-Mozaz, S., 2019. Impact of fullerenes in the bioaccumulation and biotransformation of venlafaxine, diuron and triclosan in river biofilms. Environ. Res. 169, 377–386. https://doi.org/https://doi.org/10.1016/j.envres.2018.11.036

Solórzano, L., 1969. Determination of ammonia in natural seawater by the phenol-hypochlorite method. Limnol. Oceanogr. 14, 799–801. https://doi.org/10.4319/lo.1969.14.5.0799

Sutherland, D.L., Turnbull, M.H., Craggs, R.J., 2014. Increased pond depth improves algal productivity and nutrient removal in wastewater treatment high rate algal ponds. Water Res. 53, 271–281. https://doi.org/http://dx.doi.org/10.1016/j.watres.2014.01.025

Taheran, M., Brar, S.K., Verma, M., Surampalli, R.Y., Zhang, T.C., Valero, J.R., 2016. Membrane processes for removal of pharmaceutically active compounds (PhACs) from water and wastewaters. Sci. Total Environ. 547, 60–77. https://doi.org/10.1016/j.scitotenv.2015.12.139

Thomas, D.G., Minj, N., Mohan, N., Rao, H., 2016. Cultivation of Microalgae in Domestic Wastewater for Biofuel Applications – An Upstream Approach. J . Algal Biomass Utln 7, 62–70.

Tong, L., Pérez, S., Gonçalves, C., Alpendurada, F., Wang, Y., Barceló, D., 2011. Kinetic and mechanistic studies of the photolysis of metronidazole in simulated aqueous environmental matrices using a mass spectrometric approach. Anal. Bioanal. Chem. 399, 421–428. https://doi.org/10.1007/s00216-010-4320-5

Trinh, T., van den Akker, B., Coleman, H.M., Stuetz, R.M., Drewes, J.E., Le-Clech, P., Khan, S.J., 2016. Seasonal variations in fate and removal of trace organic chemical contaminants while operating a full-scale membrane bioreactor. Sci. Total Environ.

https://doi.org/10.1016/j.scitotenv.2015.12.083

Uggetti, E., García, J., Álvarez, J.A., García-Galán, M.J., 2018. Start-up of a microalgae-based treatment system within the biorefinery concept: from wastewater to bioproducts. Water Sci. Technol. 78, 114–124.

Union, E., 1991. Council Directive 91/271/EEC of 21 May 1991 concerning urban waste-water treatment.

Van Den Hende, S., Beelen, V., Julien, L., Lefoulon, A., Vanhoucke, T., Coolsaet, C., Sonnenholzner, S., Vervaeren, H., Rousseau, D.P.L., 2016. Technical potential of microalgal bacterial floc raceway ponds treating food-industry effluents while producing microalgal bacterial biomass: An outdoor pilot-scale study. Bioresour. Technol. 218, 969–979. https://doi.org/http://dx.doi.org/10.1016/j.biortech.2016.07.065

Vieno, N., Tuhkanen, T., Kronberg, L., 2007. Elimination of pharmaceuticals in sewage treatment plants in Finland. Water Res. https://doi.org/10.1016/j.watres.2006.12.017

Villar-Navarro, E., Baena-Nogueras, R.M., Paniw, M., Perales, J.A., Lara-Martín, P.A., 2018. Removal of pharmaceuticals in urban wastewater: High rate algae pond (HRAP) based technologies as an alternative to activated sludge based processes. Water Res. 139, 19–29. https://doi.org/10.1016/j.watres.2018.03.072

Xia, S., Jia, R., Feng, F., Xie, K., Li, H., Jing, D., Xu, X., 2012. Effect of solids retention time on antibiotics removal performance and microbial communities in an A/O-MBR process. Bioresour. Technol. https://doi.org/10.1016/j.biortech.2011.11.112

Xiong, J.Q., Kurade, M.B., Abou-Shanab, R.A.I., Ji, M.K., Choi, J., Kim, J.O., Jeon, B.H., 2016. Biodegradation of carbamazepine using freshwater microalgae Chlamydomonas mexicana and Scenedesmus obliquus and the determination of its metabolic fate. Bioresour. Technol. https://doi.org/10.1016/j.biortech.2016.01.038

Young, P., Taylor, M., Fallowfield, H.J., 2017. Mini-review: high rate algal ponds, flexible systems for sustainable wastewater treatment. World J. Microbiol. Biotechnol. 33, 117. https://doi.org/10.1007/s11274-017-2282-x

Zessel, K., Mohring, S., Hamscher, G., Kietzmann, M., Stahl, J., 2014. Biocompatibility and antibacterial activity of photolytic products of sulfonamides. Chemosphere. https://doi.org/10.1016/j.chemosphere.2013.11.038

Zhang, Y., Geißen, S.U., Gal, C., 2008. Carbamazepine and diclofenac: Removal in wastewater treatment plants and occurrence in water bodies. Chemosphere. https://doi.org/10.1016/j.chemosphere.2008.07.086